\documentclass[11pt]{article}
 \usepackage{axodraw}
\usepackage{amsmath,amssymb}
\begin{document}

\begin{titlepage}
\hfill QMUL-PH-11-14

\hfill DAMTP-2011-99 

\hfill KCL-MTH-11-21

\vspace{0.25cm}
\begin{center}

{{\LARGE  \bf 	Duality Invariant Actions and Generalised Geometry}} \\

\vskip 1.5cm {David S. Berman$^\star$\footnote
{D.S.Berman@qmul.ac.uk}, Hadi Godazgar$^\dag$\footnote
{H.M.Godazgar@damtp.cam.ac.uk}, Malcolm J. Perry$^{\dag}$\footnote
{M.J.Perry@damtp.cam.ac.uk} and Peter West$^\ddag$\footnote
{Peter.West@kcl.ac.uk}}
\\
{\vskip 0.2cm
$^\star$Department of Physics,\\
Queen Mary University of London,\\
Mile End Road,
London,\\
E1 4NS, UK\\
}
{\vskip 0.2cm
$^\dag$DAMTP, Centre for Mathematical Sciences,\\
University of Cambridge,\\
Wilberforce Road, Cambridge, \\ CB3 0WA, UK\\}
{\vskip 0.2cm
$^\ddag$Department of Mathematics,\\
King's College, London\\
WC2R 2LS, UK\\}
\end{center}
\vskip 1 cm

\begin{center}
\today
\end{center}

\noindent

\vskip 1cm

\begin{abstract}

\noindent We construct the non-linear realisation of the semi-direct product of $E_{11}$ and its first fundamental representation at lowest order and appropriate to spacetime dimensions four to seven. This leads to
a non-linear realisation of the duality groups and introduces fields that
depend on a generalised space which possess a generalised vielbein. We
focus on the part of the generalised space on which the duality groups
alone act and construct an invariant action.
\end{abstract}

\end{titlepage}

\section{Introduction}
\label{1}

Nobody really knows what M-theory is, although quite a lot is known about its
various limits. These include the five ten-dimensional string theories, along with eleven-dimensional supergravity which describes the low energy effective
action of the IIA string at strong coupling. In fact the low energy effective
actions of the different string theories given by their respective supergravities contain both nonperturbative and perturbative information. As such, the U-duality web relating these theories can be tested in detail using the supergravity description. Common to all these theories is a notion of spacetime described either by a vielbein or a metric
together with various gauge fields and fermions which propagate in the
spacetime. It seems strange that in a theory that is supposed to unify
the forces of nature, one treats the gravitational field geometrically
whereas others are painted on to the geometrical spacetime. Our aim here
is to develop a more democratic approach. 

Such an approach was advocated in \cite{w42} where it was conjectured that the non-linear realisation of a certain Kac-Moody
algebra called $E_{11}$ is an extension of  eleven dimensional
supergravity. In \cite{w42}, spacetime is not encoded in an $E_{11}$
covariant way. Spacetime can be introduced by considering the  non-linear
realisation of the semi-direct product of
$E_{11}$ with its a fundamental  representation, usually called the first fundamental representation \cite{w31}.  This
semi-direct product is  explained in detail later. Semi-direct product constructions are well known, for example, the Poincar\'{e} group  is just the semi-direct product of the Lorentz
group and its vector representation, that is the  spacetime
translations. The first fundamental  representation contains as its
first component the spacetime translations, then a  two  and five form
as well as  an infinite number of other  objects.  There is considerable
evidence to suggest that  all  brane charges are contained in this
representation \cite{w31, w29, w25, w9} and for each field  in the
$E_{11}$ part of the non-linear realisation, there is a corresponding
element in this representation \cite{w29}. The   inclusion of
the  first fundamental representation in the non-linear realisation leads
to  a generalised spacetime with a coordinate for every brane charge
and  for every field.  Thus for the metric we find  the usual coordinates $x^a$ of spacetime, for
the three form new coordinates $x_{a_1a_2}$ and for the six form new coordinates $x_{a_1\ldots a_6}$
and so on
\cite{w31}. The $E_{11}$ part of the  formulation is also democratic in the
sense that
$E_{11}$ contains all the duality symmetries together with  all the
corresponding fields \cite{w15}.

To understand better this development, it is useful to recall some of the background.
In the early days of particle physics, with the recognition of the importance of symmetries, non-linear realisations played an important role. In particular, Goldstone's theorem states that if a rigid symmetry
$G$ is spontaneously broken to a subgroup $H,$ then there are $( \textup{dim} G -  \textup{dim} H)$ massless particles. Furthermore it was realised that the dynamics of
these particles is controlled by the non-linear realisation of $G$ with
local subgroup $H$. In the case of the chiral symmetry, the group $G$ is $SU(2)\otimes SU(2)$,
the  subgroup $H$  is the diagonal subgroup $SU(2)$ and  the three
massless particles are the three pions in the limit of zero mass. The
dynamics of the pions can be accounted for by this non-linear
realisation  \cite{weinberg, schwinger2, schwinger1, nlsigma, GL}. The general formulation of such non-linear realisation for any
group is given in references \cite{WessZumino, CWZ, CCWZ}.

Of course it was only later that the importance of gauge symmetries was
understood, and it was realised that pions were made of quarks subject to forces
controlled by an SU(3) gauge theory. However, this only serves to
illustrate that in the context of spontaneously broken symmetries,
non-linear realisations provide a way of finding the underlying symmetry
even though the fundamental degrees of freedom are not known.

The non-linear realisations used in the early days of particle physics,
and just discussed above,  are essentially a coset construction of $G$ with
respect to
$H$ and spacetime is a dummy variable as far as group theory is
concerned.  The sigma model usually describes this coset construction. However,
one can also construct non-linear realisations in which the group contains
generators associated with spacetime and in particular the spacetime
translations.  For
these non-linear realisations spacetime arises naturally as it
parametrises the part of the group element that includes the generators
associated with spacetime. One early paper using this method was
\cite{ISS} where the non-linear realisation with
$G= $ GL$(4,\mathbb{R})$ and $H=O(3,1)$ was studied in the context of
general relativity. However, it was  Borisov and Ogievetsky \cite{BO} who
showed that general relativity in four dimensions could be reformulated as
a non-linear realisation of the groups $G= $ GL$(4,\mathbb{R}) \ltimes I^4$ and $H=O(3,1)$.  Here GL$(4,\mathbb{R}) \ltimes I^4$ is the semi-direct
product of the groups GL$(4,\mathbb{R})$ and the group
$I^4$ of  spacetime translation generators. It is the inclusion of
the latter that lead to the presence  of the spacetime coordinates in the
theory.  In fact the dynamics of this non-linear realisation was only
unique up to a few constants and these were fixed to precisely the right
values if one demanded that the theory be also invariant under the
conformal group, also non-linearly realised. Another use of such
non-linear realisations was by  Volkov and Akulov \cite{VA} who used it
to compute the dynamics of  the massless fermion that results from  the
breaking of supersymmetry and postulated that it could be a neutrino.

The $E_{11}$ conjecture arises from the recognition that the eleven
dimensional supergravity theory is a non-linear realisation and that this
leads to an algebra including the spacetime
translations \cite{w45}. When the spacetime translations are
omitted from the algebra, it can be extended to a Kac-Moody algebra and the smallest such algebra is
$E_{11}$ \cite{w42}. As the
non-linear realisation involves the spacetime generators,  it
cannot be a sigma model.  To include the spacetime translations in a
covariant manner the first fundamental representation of $E_{11}$,
denoted by $l_1$,  is considered. This is the smallest $E_{11}$
representation that contains spacetime translations. The original and early papers
introduce the spacetime generators by hand and so only included the
first component of the $l_1$ representation.

An earlier work that
formulates the gauge fields of the maximal  supergravity theories as a
non-linear realisation using a graded algebra is  \cite{cremmerjulialupope}. The non-linear realisation in \cite{cremmerjulialupope} does not contain any spacetime generators.

\begin{figure}[http]
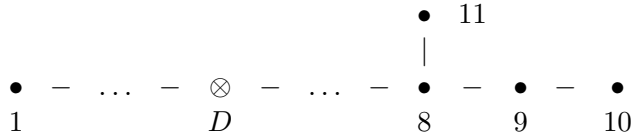

$$
\begin{array}{ccccccccccccccccc}
     & &       & &         & &       &  &\bullet&11&       & &        \\
     & &       & &       & &       &  &   |   &  &          &
&         \\
\bullet &-&\ldots&-&\otimes &-&\ldots&- &\bullet&- &\bullet&-&
\bullet \\
    1    & &         & &    D     & &      &  &   8   &  &   9   &
&    10    \\
\end{array}
$$
\caption{The $E_{11}$ Dynkin diagram with node $D$ deleted.}
\label{fig0}
\end{figure}

A theory in $d$ dimensions\footnote{In this paper $d$ corresponds to the directions in which the duality acts. In \cite{w42, w40, w25, w13, w11, w8, w7}, the complementary view is taken whereby $d$ is $11-d$ of this paper.} can be found \cite{w42, w40, w25, w13, w11, w8, w7} by taking the non-linear realisation of $E_{11} \ltimes l_1$ with the decomposition of $E_{11}$ into the subalgebra $GL(D)\otimes E_{d},$ where $D=11-d.$ This can be done by deleting node $D$ in the
$E_{11}$ Dynkin diagram in figure \ref{fig0}.

The $E_{d}$ factor in the subalgebra
$GL(D)\otimes E_{d}$ is the well known $E_{d}$
symmetry\footnote{Throughout this paper, we are considering the split forms of the exceptional groups, usually denoted $E_{d(d)}.$}  \cite{so(8), Julia, TM} which has been  known to be a symmetry of the maximal
supergravity theory in $D$ dimensions for many years. Thus these
symmetries  naturally emerge.   The
$GL(D)$ factor in the subalgebra, together with the spacetime
translations in $D$ dimensions which are contained in the $l_1$
representation give rise to gravity in $D$ dimensions as they
should according to \cite{BO}. Indeed this confirms
that we have found a  theory in $D$
dimensions. In the decomposition of
$E_{11}$ into $E_{d}$ one finds the expected fields of $D$ dimensional
supergravity as well as a hierarchy of form fields \cite{w13, BDN}, which play an important role in gauged supergravities, as
well as an infinite number of higher level fields.  The $l_1$
representation is also decomposed into representations of
$GL(D)\otimes E_{d}$ and in addition to the spacetime translations in
$D$ dimensions one finds an infinite number of coordinates beginning with
some  coordinates, which are scalars under GL$(D)$ but transform under
$E_{d}$  indeed in $d=4,5,6,7,8$ dimensions they  belong to  the 10,
$ \overline {16}$,
$\overline  {27}$, 56 and $248\oplus 1$ representations of SL(5), SO(5,5),
$E_6$,
$E_7$ and $E_8$ respectively \cite{w25, w23}.
The non-linear realisation of $E_{11} \ltimes l_1$ not only gives rise to generalised spacetime, but it also leads to a generalised vielbein
which is determined in terms of the $E_{11}$ fields and depends on the
generalised spacetime. In this paper, the theory in $d$ dimensions is considered. We explicitly construct the generalised vielbeins and the corresponding dynamics.

For future reference, in table \ref{table1} we recall the U-duality groups in the
various dimensions.

\begin{table}[htbp]
\centering
\begin{tabular}{ | c | c | c | c | } 
\hline
$D$ & $d$ & $G$ & $H$ \\ \hline
3 & 8 & $E_{8}$ & SO(16) \\
4 & 7 & $E_{7}$ & SU(8) \\
5 & 6 & $E_{6}$ & USp(8) \\
6 & 5 & SO(5,5) & SO(5)$\times$SO(5) \\
7 & 4 & SL(5) & SO(5) \\
8 & 3 & SL(3)$\times$SL(2) & SO(3)$\times$SO(2) \\
9 & 2 & SL(2) & SO(2) \\
10 & 1 & SO(1,1) & 1 \\ \hline
\end{tabular}
\caption{The duality groups that appear on the reduction of 11-dimensional supergravity to $D-$dimensions.}
\label{table1}
\end{table}

In fact one can formulate the dynamics of strings, membranes etc in the presence of the
background fields  as an
$E_{11} \ltimes l_1$ non-linear realisation \cite{w23}. The difference
compared to the non-linear realisation used to construct the supergravity
theories was in the choice of  local subalgebra. In \cite{w23} the
coordinates of the generalised spacetime specifies   the dynamics
of the brane.

An enlarged spacetime also appeared in the context of the first
quantised string \cite{Tseytlin, Tseytlinb, Duff} and membrane \cite{DuffLu} where the usual
spacetime is extended to include additional coordinates describing string winding modes. The aim in
the case of the string is to make the T-duality symmetry manifest by
introducing additional coordinates corresponding to string winding modes.
This is then extended to the membrane in \cite{DuffLu}, where new coordinates are
introduced corresponding to membrane windings, so that the U-duality
group is made manifest.
The work in \cite{DuffLu} is further developed in \cite{davidmalcolm} for
the SL(5) duality group to a give duality-invariant dynamics
for fields living on a space whose coordinates belong to
the ten dimensional representation of SL(5). The invariant dynamics is constructed using a generalised metric given in terms of the background supergravity fields, and later extended to the
duality group SO(5,5) in \cite{BGP}. The usual way in which duality
groups appear is where one dimensionally reduces eleven dimensional
supergravity. The duality group then acts on the
components of the fields in the Kaluza-Klein directions. In \cite{davidmalcolm, BGP}
the opposite approach is taken; the duality group acts on the space where the
fields have spacetime dependence, i.e.\ no Killing directions are assumed.

In \cite{hillmann}, the non-linear realisation of $E_{11} \ltimes l_1$ decomposed
to
$GL(4)\otimes E_7$ is constructed. The part of the $l_1$ representation that is kept leads to
the usual coordinates of the four  dimensional spacetime and  also the
coordinates which are scalars under GL(4) but transform as a 56
dimensional representation of
$E_7$.  The non-linear realisation is then used in \cite{hillmann} to construct an invariant action.

In the present paper, we will show how the results of
\cite{davidmalcolm, BGP} can be derived in a very
straightforward way from  the
$E_{11} \ltimes l_1 $ non-linear realisation discussed above.
Indeed we construct  the non-linear realisation $E_{11} \ltimes l_1$
decomposed to
$GL(D)\otimes E_{d}$ suitable to $d$ dimensions. We restrict the
$l_1$ representation to contain only  the coordinates  that
are scalars under GL$(D)$, which turn out to
transform as the  $ \overline{10}$, 16, 27 and 56 dimensional representations of
SL(5), SO(5,5),
$E_6$  and
$E_7$ respectively. We construct invariant
actions where the fields are defined on these generalised spacetimes.

In section \ref{2}, we revisit four dimensions and  the SL(5) duality
group. The generalised metric in this case was constructed in
\cite{davidmalcolm} using M2-brane considerations. In section \ref{2},
the non-linear realisation of the SL(5) motion group is used to construct
the generalised metric, which is the same as in \cite{davidmalcolm} up to
a conformal factor. Then, we give a review of $E_{11}$ and its first
fundamental representation in section \ref{3}. In this section, we also
review the non-linear realisation of $E_{11} \ltimes l_1$. In section
\ref{4}, the familiar example of four dimensions is used to illustrate
how the non-linear realisation of
$E_{11} \ltimes l_{1}$ can be used to find the generalised metric and the
dynamics. Then, in sections \ref{5}, \ref{6} and \ref{7}, we proceed to
carry the same procedure in five, six and seven dimensions. In each case
we find the generalised metric and formulate the dynamics in terms of this
object to give a duality invariant action that reproduces the usual
11-dimensional supergravity action.

\section{SL(5) generalised metric}
\label{2}

In this section we consider in detail the duality group SL(5) and
give a rather pedestrian presentation. This will allow us to study
the SL(5) duality group and the ten dimensional spacetime that
occurs in this case in isolation. We will just present the algebra
rather than derive it from $E_{11} \ltimes l_1$, we will explain in
detail the way the non-linear realisation leads to the generalised
metric and the corresponding dynamics. This will allow one to gain
some understanding of the technical aspects of the non-linear
realisations used without all the complications of the $E_{11}
\ltimes l_1$ algebra.

The starting point of the non-linear realisation method is the
duality group, from which we form the corresponding motion group. The
semi-direct product of a group with a representation of the group
defines the motion group \cite{wigner, mackey}. For example, the
Poincar\'{e} group is the motion group of the Lorentz group. The SL(5) algebra itself is given by 24 tracefree generators. In the
fundamental representation of SL(5), the generators can be chosen to be
$$ \left(M^{I}_{\;\;J} \right)^{P}_{\;\;Q} = - \delta^{P}_{J} \delta^{I}_{Q} + {{1}\over{5}} \delta^{I}_{J} \delta^{P}_{Q}. $$
The indices $I, J=1, \dots 5$ and are the generator labels, while $P, Q$ are matrix indices which also run from 1 to 5 because we are in
the fundamental representation. It can be explicitly checked that the
generators satisfy the expected SL(5) commutation relations
\begin{equation}
[M^{I}_{\;\;J}, M^{K}_{\;\;L} ] = \delta^{K}_{J} M^{I}_{\;\;L} -
\delta^{I}_{L} M^{K}_{\;\;J}.
\label{(1)}
\end{equation}

We will construct the motion group of SL(5) where the translation
generators form a ten-dimensional representation. This is similar to
the construction of the Poincar\'{e} group from the Lorentz group.
The translation generators form the 10 of SL(5), which we denote by
$P_{IJ},$ where the indices again run from 1 to 5 and $P$ is
antisymmetric in these indices so that we have ten generators.

The translation generators all commute with each other, and their
commutation relations with the group generators and the translation
generators are
\begin{equation}
[M^{I}_{\;\;J}, P_{KL} ] = - 2 \, \delta^{I}_{[K} P_{|J|L]} + {{2} \over{5}} \delta^{I}_{J} P_{KL}.
\label{(2)}
\end{equation}
The coefficient of the first term on the right-hand side is fixed by
the Jacobi identities, while the coefficient of the second is
determined by the requirement that the generators $M$ are tracefree.

The SL(5) duality group first appeared when a Kaluza-Klein reduction of
eleven-dimensional supergravity was made on a flat 4-torus. In our
picture, SL(5) appears as a group which controls the geometry of the
4-manifold itself. Unlike Kaluza-Klein reduction, the 4-manifold is not
associated with any Killing vectors. The fields \textit{depend} on the
coordinates in the directions of the 4-manifold. We will ignore the
dependence of all fields on directions orthogonal to the 4-manifold. This is \textit{opposite} to Kaluza-Klein reduction. Thus, if we were considering eleven-dimensional supergravity there
will be seven directions that are ignored. However, as was found in
\cite{davidmalcolm}, the four directions must be augmented by
the six winding directions associated with the M2 branes charges. There a total of ten dimensions of the extended space associated with the four physical spatial directions. The ultimate
interpretation of these extra dimensions is presently a little
unclear but is discussed in \cite{BGGP}, where the local symmetries of M-theory is explored in the context of generalised geometry. This approach leads to the physical section condition for M-theory generalised geometry. The extra dimensions are M-theoretic generalisations of the winding coordinates found in doubled field theory \cite{dft1, dft2, dft3, dft4}.

To make the relation to the usual fields and coordinates clear, we
will decompose the SL(5) group into its SL(4) $\times$ U(1) subgroup.
The SL(4) corresponds to the usual four spatial directions. We let
\begin{equation}
M^{I}_{\;\;J} = \begin{cases}
               M^{i}_{\;\;j} \\
         M^{5}_{\;\;j} = {{1}\over{6}} \epsilon_{jklm} R^{klm} \\
         M^{i}_{\;\;5} = {{1}\over{6}} \epsilon^{iklm} R_{klm}
              \end{cases}.
\end{equation}
The indices labelled by $i,j, \dots $ are GL(4) indices that run from
1 to 4. Note that $$M^{5}_{\;\;5}= - \sum_{i=1}^{4} M^{i}_{\;\;i},$$ by the
tracelessness of $M^{I}_{\;\;J}.$ The generator $\sum M^{i}_{\;\;i}$
which we will denote by $M$, gives the scaling of generators in the GL(4) decomposition and so determines their U(1) charge. The generator
$M^{i}_{\;\;j}$ can be shifted by $M,$ and indeed we will shift
\begin{equation}
M^{i}_{\;\;j} \rightarrow K^{i}_{\;\;j} = M^{i}_{\;\;j} - \delta^{i}_{j} M.
\label{(4)}
\end{equation}
The dilatation is now given by $$ K \equiv \sum K^{i}_{\;\;i} = -3 M, $$
which generates the U(1) subgroup of SL(5). With this choice,
\[
[K, R^{klm} ] = 3 \, R^{klm}, \qquad  [K, R_{klm} ] = - 3 \, R_{klm} \qquad \textup{and} \qquad  [K, K^{i}{}_{j} ] = 0,
\]
so that $K$ counts the index of the GL(4) representations, in other
words, its U(1) charge. Other choices can of course be made, but
these will result in more complicated commutation relations between
the $K^{i}{}_{j}$ generator and generalised translation generators.

We can now rewrite the SL(5) algebra in terms of the GL(4) and U(1)
generators
\begin{gather}
[ K^{i}_{\;j} , K^{l}_{\;m} ] = \delta^{l}_{j} K^{i}_{\;m} - \delta^{i}_{m} K^{l}_{\;j}, \qquad
[ R^{i_1 \dots i_3} , R_{j_1 \dots j_3} ] = 18 \, \delta^{[i_1 i_2}_{[j_1 j_2} K^{i_3]}_{\;j_3]} - 2 \, \delta^{i_1 \dots i_3}_{j_1 \dots j_3} K, \\
[ K^{i}_{\;j} , R_{k_1 \dots k_3} ] = -3 \, \delta^{i}_{[k_1} R_{k_2 k_3]j},  \qquad
[ K^{i}_{\;j} , R^{k_1 \dots k_3} ] = 3 \, \delta^{[k_1}_{j} R^{k_2 k_3]i};
\end{gather}
all other commutators vanish. The fully antisymmetrised Kronecker
delta function is defined to be
\begin{multline*}
\delta^{i_1 \dots i_p}_{j_1 \dots j_p} = \delta^{[i_1}_{j_1} \dots \delta^{i_p]}_{j_p} = {{1}\over{p!}} \left(\delta^{i_1}_{j_1} \dots \delta^{i_p}_{j_p} + (\textup{all remaining even permutations of }i_1
\dots i_p) \right. \\ 
- (\textup{all odd permutations of }i_1
\dots i_p) \Big),
\end{multline*}
making a total of $p!$ terms in the parentheses.

Now that we have the SL(5) algebra,  we similarly write the translation
generators
\begin{equation}
P_{I J} = \begin{cases}
               P_{i 5} = P_{i} \\
         P_{i j} = {{1}\over{2}} \epsilon_{ijkl} Z^{kl}
              \end{cases}.
\label{(7)}
\end{equation}
The 10-dimensional representation in  terms of a GL(4) decomposition is
made in order to relate the translation generators to the generators of
ordinary spatial translations in four-dimensions $P_{i},$ together with
the generalised translations $Z^{ij},$ which correspond to windings of
the M2-brane.

Now, from equation \eqref{(2)},  the rest of the commutation
relations of
the algebra are
\begin{gather}
[ K^{i}_{\;j} , P_{k} ] = - \delta^{i}_{k} P_{j} - {{1}\over{5}}
\delta^{i}_{j} P_{k}, \qquad
[ K^{i}_{\;j} , Z^{kl} ] = 2 \, \delta^{[k}_{j} Z^{|i|l]} -  {{1}\over
{5}} \delta^{i}_{j} Z^{kl}, \label{(8)} \\
[ R_{ijk} , P_{l} ] = 0, \quad
[ R_{ijk} , Z^{mn} ] = 3! \, \delta^{mn}_{[ij} P_{k]}, \quad
[ R^{ijk} , P_{l} ] = 3 \, \delta^{[i}_{l} Z^{jk]}, \quad
[ R^{ijk} , Z^{mn} ] = 0. \label{(9)}
\end{gather}
Note that for the translation generators the U(1) generator $K$ does
not count the index of the generator as it did for the SL(5) generators;
$$[K, P_{i}]= {9\over5} P_{i} \qquad \textup{and} \qquad [K, Z^{ij}]= {11\over5} Z^{ij}.$$

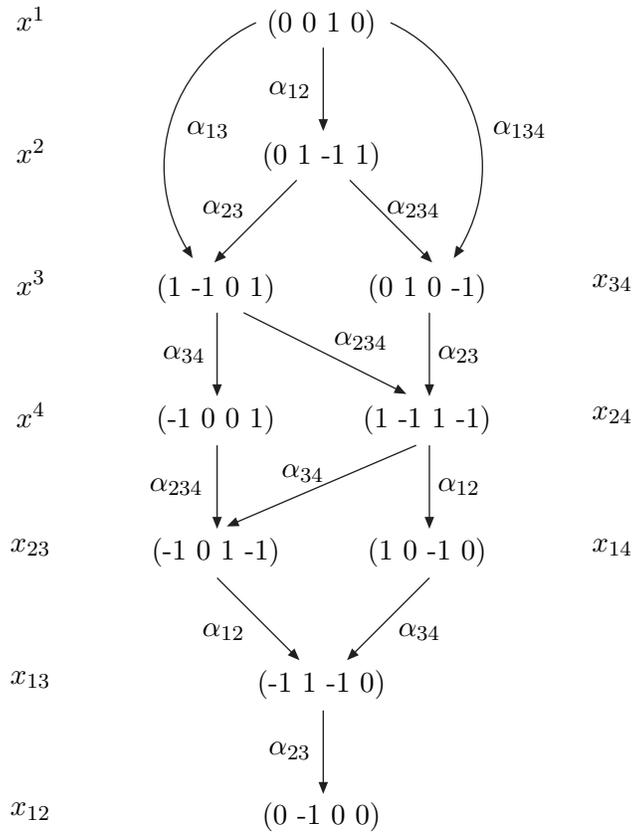
\begin{figure}
\begin{center} \begin{picture}(320,330)(0,0)
\Text(160,320)[c]{(0 0 1 0)}
\LongArrow(160,310)(160,280)  \Text(140,295)[l]{$\alpha_{12}$}
\Text(160,270)[c]{(0 1 -1 1)}
\LongArrow(150,260)(120,230)  \Text(115,250)[l]{$\alpha_{23}$}
\LongArrow(170,260)(200,230)  \Text(205,250)[r]{$\alpha_{234}$}
\Text(120,220)[c]{(1 -1 0 1)} \Text(200,220)[c]{(0 1 0 -1)}
\LongArrow(120,210)(120,180)  \Text(100,195)[l]{$\alpha_{34}$}
\LongArrow(200,210)(200,180)  \Text(220,195)[r]{$\alpha_{23}$}
\LongArrow(130,210)(190,180)  \Text(185,200)[r]{$\alpha_{234}$}
\Text(120,170)[c]{(-1 0 0 1)} \Text(200,170)[c]{(1 -1 1 -1)}
\LongArrow(120,160)(120,130)  \Text(95,145)[l]{$\alpha_{234}$}
\LongArrow(200,160)(200,130)  \Text(220,145)[r]{$\alpha_{12}$}
\LongArrow(195,160)(125,130)  \Text(145,150)[l]{$\alpha_{34}$}
\Text(120,120)[c]{(-1 0 1 -1)} \Text(200,120)[c]{(1 0 -1 0)}
\LongArrow(120,110)(150,80)   \Text(115,90)[l]{$\alpha_{12}$}
\LongArrow(200,110)(170,80)   \Text(205,90)[r]{$\alpha_{34}$}
\Text(160,70)[c]{(-1 1 -1 0)}
\LongArrow(160,60)(160,30)   \Text(140,45)[l]{$\alpha_{23}$}
\Text(160,20)[c]{(0 -1 0 0)}
\Text(50,322)[c]{$x^1$}
\Text(50,272)[c]{$x^2$}
\Text(50,222)[c]{$x^3$}
\Text(50,172)[c]{$x^4$}
\Text(50,122)[c]{$x_{23}$}
\Text(50,72)[c]{$x_{13}$}
\Text(50,22)[c]{$x_{12}$}
\Text(270,222)[c]{$x_{34}$}
\Text(270,172)[c]{$x_{24}$}
\Text(270,122)[c]{$x_{14}$}
\LongArrowArc(160,265)(60,115,215) \Text(125,280)[r]{$\alpha_{13}$}
\LongArrowArcn(160,265)(60,65,325) \Text(245,280)[r]{$\alpha_{134}$}
\end{picture} \end{center} 
\caption{The weight diagram of the 10-dimensional representation of SL(5).} \label{fig:weight} \end{figure}

In figure \ref{fig:weight}, the weight diagram of the ten-dimensional representation of SL(5) is presented. The weight diagram is generated by subtracting positive roots from the weights (equivalently adding negative roots to the weights). The generators 
$$K^{i}{}_{j}, R^{k_1 \dots k_3}, R_{k_1 \dots k_3}$$
are associated to the roots of SL(5)
$$\alpha_{ij},  \alpha_{k_1 \dots k_3}, -\alpha_{k_1 \dots k_3}.$$ 
The root lattice is generated by adding arbitrary multiples of positive roots to these. For example, $$\alpha_{12} +\alpha_{23} = \alpha_{13} \qquad {\rm and } \qquad \alpha_{12} +\alpha_{234} = \alpha_{134},$$
from which the commutators 
$$[K^{1}{}_{2},K^{2}{}_{3}]=K^{1}{}_{3} \qquad {\rm and } \qquad [K^{1}{}_{2},R^{234}]=R^{134}$$ 
can be constructed.
Similarly, the translation generators $P_{i}$ and $Z^{ij}$ are associated to the weights labelled by $x^{i}$ and $x_{ij}$ in figure \ref{fig:weight}. The $x^{i}$ and $x_{ij}$ then become coordinates of the extended space. The commutation relations of the motion group of SL(5) encode how the roots act on the weights. The negative roots 
$$\alpha_{ij}, \quad {\rm for} \; i<j, \qquad {\rm and} \qquad  \alpha_{k_1 \dots k_3} $$ 
act on the 10-dimensional weight diagram by lowering the weights, while the positive roots  
$$\alpha_{ij}, \quad {\rm for} \; i>j, \qquad {\rm and} \qquad -\alpha_{k_1 \dots k_3}$$ 
raise the weights. In figure \ref{fig:weight}, for example, $\alpha_{23}$ acts on the weight $x_{34}$ to give $x_{24}$. In terms of a commutation relation, this is $$[K^{2}{}_{3}, Z^{34}]=Z^{24},$$ which is consistent with the second equation in \eqref{(8)}.

We need to find the normalisation of the translation generators,  which
set the conventions for the tangent space metric. Let\footnote{Our
treatment of the normalisation of generators in this section is
motivated
purely by convenience. A more rigorous treatment involves the definition
of the Cartan involution of $P$ and is described in appendix \ref{A}.}
$$ \textup{tr}(P_{IJ} P_{KL}) = 2 \, \delta_{IJ, KL} = ( \delta_{IK}
\delta_{JL} - \delta_{IL} \delta_{JK} ),$$ and by inserting the
translation generators given in equation \eqref{(7)} we find that
\begin{equation}
\textup{tr}( P_{i} P_{j} ) = \delta_{ij}, \qquad  \textup{tr}( Z^{ij}
Z^{kl} ) = 2 \, \delta^{ij, kl}, \qquad \textup{tr}( P_{i} Z^{kl} ) = 0,
\end{equation}
where $$\delta^{ij,kl}= {{1}\over{2}} ( \delta^{ik} \delta^{jl} - \delta^{il} \delta^{jk} ).$$

The generalised metric is constructed using the non-linear realisation
method. We start by writing the group element
$$g_{l} = \textup{e}^{x^{i} P_{i}} \textup{e}^{{{1}\over{\sqrt{2}}} x_
{kl} Z^{kl}},$$
where $x^{i}$ are the conventional coordinates, and $x_{kl}$
are the ``winding coordinates.''
The coefficient of the each exponent is such that the tangent space
metric takes the canonical form, i.e.\ \begin{equation}
\textup{tr}(g^{-1}_{l} \textup{d}g_{l} g^{-1}_{l} \textup{d}g_{l}) =
\delta_{ij} \textup{d}x^{i} \textup{d}x^{j} + \delta_{kl,mn} \textup
{d}x_{kl} \textup{d}x_{mn}.
\label{(11)}
\end{equation}
The group element that defines the fields is
$$g_{E} = \textup{e}^{h_{i}^{\;j} K^{i}_{\;j}} \textup{e}^{{{1}\over
{3!}} C_{ijk} R^{ijk}}.$$
$C_{ijk}$ is the 3-form potential of M-theory restricted to the 4-space and $h_{i}{}^{j}$ determines the vielbein.

The generalised vielbein, $E,$ is given by the Maurer-Cartan form of $g_{l}$ conjugated by $g_{E}$
\begin{equation}
L_{A} E_{\Pi}{}^{A} \textup{d}z^{\Pi}= g^{-1}_{E} g^{-1}_{l} \textup
{d}g_{l} g_{E},
\label{(12)}
\end{equation}
where $ L_{A}= (P_{i}, Z^{kl}/ \sqrt{2})$ and $\textup{d}z^{\Pi} =
(\textup{d}x^{\mu}, \textup{d} x_{\mu \nu}).$ Latin letters indicate
tangent space indices, while Greek letters label spacetime indices. The
normalisation of $L_{A}$ has been arranged so that $\textup{tr}( L_{A}
L_{B})=\delta_{A B}.$ In terms of the generalised vielbein, the
generalised line element is given by
$$\textup{Tr}(L_{A} E_{\Pi}{}^{A} \textup{d}z^{\Pi}L_{B} E_{\Sigma}{}^
{B} \textup{d}z^{\Sigma})= E_{\Pi}{}^{A} E_{\Sigma}{}^{B} \delta_{AB}
\textup{d}z^{\Pi}  \textup{d}z^{\Sigma}.$$
Consequently, the generalised metric is
\begin{equation}
M_{\Pi \Sigma} = E_{\Pi}{}^{A} E_{\Sigma}{}^{B} \delta_{AB}.
\label{(13)}
\end{equation}
One can regard the 1-forms $E_{\Pi}{}^{A} \textup{d}z^{\Pi}$ as an
orthonormal basis in our generalised tangent space.

The Cartan metric of $g_{l}$ gives the generalised tangent space
metric,
equation \eqref{(11)}. It can be thought of as the generalised metric of
flat space with vanishing 3-form potential. Conjugating the Maurer-Cartan
form by $\textup{e}^{h_{i}^{\;j} K^{i}_{\;j}}$ gives the vielbein for
curved space and further conjugation by  $\textup{e}^{{{1}\over{3!}}
C_{ijk} R^{ijk}}$ gives the dependence of the generalised vielbein on
the
3-form potential.

We now find the result of conjugating $g_{l}^{-1} \textup{d}  g_{l}$ by
the group element corresponding to the $K$ generator. The Maurer-Cartan
form of $g_{l}$ is
\begin{equation}
g_{l}^{-1} \textup{d}g_{l} = \textup{d}x^{i} P_{i} + {{1}\over{\sqrt
{2}}} \textup{d} x_{kl} Z^{kl}.
\label{(14)}
\end{equation}
Using the Hadamard formula\footnote{The adjoint map ad is defined by 
$(\textup{ad}^{n} X) Y= [X[X[X \dots [X,Y]]]\dots],$ where there are
$n$ commutators \cite{fuchsschweigert}.}
\begin{equation*}
\textup{e}^{X} Y \textup{e}^{-X} = \textup{e}^{ \textup{ad} X} Y,
\end{equation*}
we can evaluate
\begin{align}
\textup{e}^{-h_{i}^{\;j} K^{i}_{\;j}} \textup{d}x^{m} P_{m} \textup{e}
^{h_{k}^{\;l} K^{k}_{\;l}} &= \textup{d}x^{k} \sum_{n=0}^{\infty}
{{(-1)^n}\over{n!}} h_{i_1}^{\;\;j_1} \dots h_{i_n}^{\;\;j_n} \big[ K^
{i_n}{}_{j_n} , [  \dots  [ K^{i_1}_{\;\;\;j_1}, P_{k} ] \dots ]
\big], \notag \\
&= \textup{d}x^{i} \sum_{n=0}^{\infty} {{1}\over{n!}} \sum_{m=0}^{n}
{{1}\over{5^m}} \begin{pmatrix} n \\ m \end{pmatrix} (\textup{tr} h)
^m (h^{n-m})_{i}^{\;\;j} P_{j}, \notag \\
&= \textup{d}x^{i} \sum_{m=0}^{\infty}  {{1}\over{5^{m} m!}} (\textup
{tr} h)^m \sum_{n=0}^{\infty} {{1}\over{n!}} (h^n)_{i}^{\;\;j} P_{j},
\notag \\
&=  \textup{det}(\textup{e}^h)^{1/5} (\textup{e}^h)_{\mu}^{\;\;j} \,
\textup{d}x^{\mu} P_{j}, \label{(15)}
\end{align}
where in going to the second line we have used the first commutation
relation in the line of equations labelled \eqref{(8)}, and in the
last equality we have used $ \textup{det}(\textup{e}^h) = \textup{e}^
{\textup{tr}(h)}.$ We can identify $\textup{e}^h$ with the vielbein
corresponding to usual spatial metric. In the last line, we have used
a Greek letter as an index on $\textup{d}x$ because a distinction
should be made between the index on the translation generator which
should be thought of as a tangent space index, and the index on the $\textup{d}x,$ which is a space index. Space is thus endowed with the
metric
\begin{equation}
g_{\mu \nu}= (\textup{e}^h)_{\mu}^{\;\;i} (\textup{e}^h)_{\nu}^{\;
\;j} \delta_{ij}.
\label{(16)}
\end{equation}

The remaining term in the Maurer-Cartan form, \eqref{(14)}, can be
conjugated by the group element of the $K$ generator in a similar
way. For the $\textup{d} x_{kl} Z^{kl}$ term we can again use the
Hadamard formula and find
\begin{align}
&\textup{e}^{-h_{i}^{\;j} K^{i}_{\;j}} \textup{d} x_{mn} Z^{mn}
\textup{e}^{h_{k}^{\;l} K^{k}_{\;l}} \notag \\
=& \sum_{n=0}^{\infty} {{(-1)^n}\over{n!}} \textup{d}
x_{mn} \left(\textup{ad}(h K)\right)^n Z^{mn}, \notag \\
=& \textup{d}x_{mn} \sum_{n=0}^{\infty} {{(-1)^n}\over{n!}} \sum_{m=0}
^{n} \sum_{p=0}^{m} \begin{pmatrix} n \\ m \end{pmatrix} \begin
{pmatrix} m \\ p \end{pmatrix} (h^p)_{i}{}^{m} (h^{m-p})_{j}{}^{n}
\left(-{{1}\over{5}} \textup{tr} h\right)^{n-m} Z^{ij}.
\label{(17)}
\end{align}
The easiest way to prove the second equality is to use induction on $n$. We then interchange the summations in equation \eqref{(17)},
taking care of the limits of the summations, to write the expression
on the right-hand side as a product of three exponentials
\begin{equation}
\textup{e}^{-h_{i}^{\;j} K^{i}_{\;j}} \textup{d}x_{mn} Z^{mn} \textup
{e}^{h_{k}^{\;l} K^{k}_{\;l}} = \textup{det}(\textup{e}^h)^{1/5}
(\textup{e}^{-h})_{i}{}^{\mu} (\textup{e}^{-h})_{j}{}^{\nu} \, \textup
{d}x_{\mu \nu} Z^{ij}.
\label{(18)}
\end{equation}
As above, the indices on the translation generators are tangent space
indices and the indices on the differential 2-form are space indices.
$(\textup{e}^{-h})_{i}^{\;\;\mu}$ is the inverse vielbein
corresponding to the metric $g$ in equation \eqref{(16)}.

We have constructed the generalised vielbein in a space with metric
$g.$ To find the dependence of the generalised vielbein, and metric,
on the 3-form potential $C,$ we will conjugate by the group element
corresponding to the $R^{ijk}$ generator. The commutation relations
of the $R^{ijk}$ generator with the translation generators are given
in equations \eqref{(9)}, from which it can be seen that $R^{ijk}$
sends the translation generators into one another---more precisely, $P$ is sent to $Z.$  The generator $R_{ijk}$ has the opposite effect.
Therefore, unlike before when conjugation by the group element
corresponding to $K$ leads to an infinite series, in this case the
sum will truncate because the commutation relation of $R^{ijk}$ and
$Z^{mn}$ vanishes. So there will only be a finite order dependence on
the 3-form potential. We begin by conjugating the term proportional
to $P_{i},$ \eqref{(15)},
\begin{align}
&\textup{e}^{-{{1}\over{3!}} C_{j_1 \dots j_3} R^{j_1 \dots j_3}}
\textup{e}^{-h_{k}{}^{l} K^{k}{}_{l}} \textup{d}x^{i} P_{i} \textup{e}
^{h_{k}{}^{l} K^{k}{}_{l}} \textup{e}^{{{1}\over{3!}} C_{j_1 \dots
j_3} R^{j_1 \dots j_3}} \notag \\
=&  \textup{det}(\textup{e}^h)^{1/5} (\textup{e}^h)_{\mu}{}^{i} \,
\textup{d}x^{\mu} \left( P_{i} - {{1}\over{3!}} C_{j_1 \dots j_3} [ R^
{j_1 \dots j_3} , P_{i}] \right. \notag \\
& \hspace{56mm} \left. + {{1}\over{2}} {{1}\over{(3!)^2}} C_{j_1
\dots j_3} C_{k_1 \dots k_3} [ R^{j_1 \dots j_3}, [R^{k_1 \dots
k_3} , P_{i}]] + \dots \right) \notag \\
=&  \textup{det}(\textup{e}^h)^{1/5} (\textup{e}^h)_{\mu}{}^{i} \,
\textup{d}x^{\mu} \left( P_{i} - {{1}\over{2}} C_{ijk} Z^{jk}
\right), \label{(19)}
\end{align}
using commutation relations \eqref{(9)}. As stressed earlier, the
series truncates.

The conjugation of the term proportional to $Z^{ij}$ is trivial
because $[R^{ijk}, Z^{mn}]=0.$
\begin{align}
g^{-1}_{h} g^{-1}_{l} \textup{d}g_{l} g_{h} = \textup{det}(\textup{e}
^h)^{1/5} (\textup{e}^h)_{\mu}{}^{i} \,\textup{d}x^{\mu} \left( P_{i}
- {{1}\over{2}} C_{ijk} Z^{jk} \right) + {{1}\over{\sqrt{2}}} \textup
{det}(\textup{e}^h)^{1/5} (\textup{e}^{-h})_{i}{}^{\mu} (\textup{e}^{-
h})_{j}{}^{\nu} \, \textup{d} x_{\mu \nu}  Z^{ij}.
\end{align}
To find the generalised vielbein we need to compare the above
expression with equation \eqref{(12)}. Hence the generalised vielbein is
\begin{equation}
E_{\Pi}{}^{A} = (\textup{det}e)^{1/5}
\begin{pmatrix}
e_{\mu}{}^{i} & -{{1}\over{\sqrt{2}}} e_{\mu}{}^{j} C_{j i_1 i_2} \\
0 & e^{\mu_1}{}_{[i_1} e^{\mu_2}{}_{i_2]}
\end{pmatrix}.
\label{(21)}
\end{equation}
Tangent space indices are written with Latin letters and Greek
letters are spatial indices. We have also abbreviated the spatial
vielbein $\textup{e}^{h}$ to $e.$ The position of the indices on $e$
indicate whether it is the spatial vielbein or inverse vielbein. If
the spatial index is lowered, i.e.\ $e_{\mu}{}^{i},$ then this is the
vielbein, and if the spatial index is raised, i.e.\ $e^{\mu}{}_{i},$
then this is the inverse vielbein.

Now from the generalised vielbein we can easily calculate the
generalised metric, using equation \eqref{(13)},
\begin{equation}
M_{KL} = g^{1/5}
\begin{pmatrix} g_{\mu \nu}+{{1}\over{2}} C_{\mu}{}^{ij} C_{\nu ij} &
-{{1}\over{\sqrt{2}}} C_{\mu}{}^{\nu_1 \nu_2 } \\ -{{1}\over{\sqrt
{2}}} C^{\mu_1 \mu_2}{}_{\nu}  & g^{\mu_1 \mu_2, \nu_1 \nu_2}
\end{pmatrix},
\label{(22)}
\end{equation}
where $g=(\textup{det}e)^{2}$ is the determinant of the metric $g_
{\mu \nu}.$ This is the same metric as in \cite{hullm, PW, davidmalcolm} except
for the factor of $g^{1/5}.$ This latter factor comes from the term
proportional to $\delta^{i}_{j}$ in the commutation relations of $[K^{i}_{\;\;j}, P_{k}]$ and $[K^{i}_{\;\;j}, Z^{kl}],$ equation \eqref{(8)}. The precise value of this coefficient was fixed by requiring
that the SL(5) generator $M^{I}_{\;\;J}$ is traceless in equation
\eqref{(2)}. It is important to note that for this particular
coefficient, i.e.\ power of $g$ multiplying the metric, we obtain a
generalised metric that does not describe the dynamical theory (see
appendix \ref{B}). In the next sections, we will consider the groups
governing generalised geometry as coming from $E_{11},$ in which case
the factor of the term proportional $\delta^{i}_{j}$ in the
commutators of $K$ and $P,Z$ is different. This results in a change
in the factor multiplying the metric to $g^{-1/2},$ rather than $g^{1/5}.$ The corresponding generalised metric can be used to construct
the dynamics and naturally incorporates the measure in precisely the
correct way.

We will now review the non-linear realisation of  $E_{11} \ltimes l_{1}$
and find the generalised metrics for the SL(5), SO(5,5), $E_{6}$ and
$E_{7}$ duality groups from the $E_{11} \ltimes l_{1}$ non-linear
realisation.

\section{A review of $E_{11}$ and its first
fundamental representation and their non-linear realisation}
\label{3}

In this section, we will  review previous  work on the original
$E_{11}$ conjecture \cite{w42}:  its application to ten
\cite{w42, w40, w38} and lower dimensions \cite{w25, w13, w11, w8, w7}; the development of $E_{11}$ as an algebra
\cite{w37, w34};  its first fundamental representation
and its relation to brane charges \cite{w31, w29, w25, w23, w9}; and finally the non-linear
realisation of
$E_{11}\ltimes l_1$ \cite{w31, w23, w11, w3, w1}.
We collect together results that are found in
different papers in a single place and we will take the opportunity to
give a user friendly presentation. Some of this review is taken from the forthcoming book \cite{Westbook}.

\begin{figure}[http]
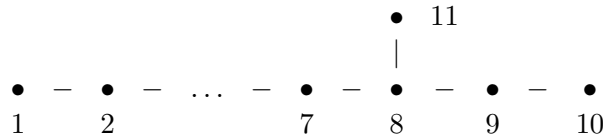

$$
\begin{array}{ccccccccccccccccc}
     & &       & &         & &       &  &\bullet&11&       & &        \\
     & &       & &       & &       &  &   |   &  &          &
&         \\
\bullet &-&\bullet&-&\ldots &-&\bullet&- &\bullet&- &\bullet&-&
\bullet \\
    1    & &   2      & &         & &   7   &  &   8   &  &   9   &
&    10    \\
\end{array}
$$
\par
\caption{The $E_{11}$ Dynkin diagram}
\label{fig2}
\end{figure}

The $E_{11}$ algebra consists of an infinite number  of generators and,
like
all Kac-Moody algebras,  it is completely determined
by its Cartan matrix, or equivalently its Dynkin diagram  given in
figure
\ref{fig2}.

Upon deleting the eleventh node of the $E_{11}$ Dynkin diagram  we find
the Dynkin diagram for SL(11). We can therefore classify the generators
of $E_{11}$ in terms of this subalgebra, or to put it another way, we
can
decompose the adjoint representation of $E_{11}$ into
representations  of
SL(11). The resulting decomposition   of $E_{11}$ can be labelled in terms of a grading usually termed the level.
Generators with non-negative levels are
given, in  increasing order, by \cite{w42, w34}
\begin{equation}
K^a{}_b (0), R^{a_1a_2a_3} (1) , R^{a_1a_2\dots a_6} (2) , R^{a_1a_2
\ldots
a_8,b}(3), \ldots,
\label{(23)}
\end{equation}
where $a,a_1,a_2, \ldots , b,\ldots =1,2,\ldots
,11$ and the number in brackets is the level of the respective
generator. The last generator
satisfies the constraint $$R^{[a_1a_2\ldots
a_8,b]}=0.$$ Of course, the sequence does not terminate, reflecting
the fact that $E_{11}$ is infinite dimensional. The level zero
generators $K^a{}_b$ obey
the  GL(11) algebra;  the  enlargement of SL(11) to GL(11) arises in
the same way as the SL(5) case in section \ref{2}. The Cartan subalgebra generator
associated with node eleven remains as part of the group even though
that node in the Dynkin diagram has been deleted. The algebra of the GL(11) generators is given by
\begin{equation}
[ K^a{}_b, K^c{}_d]=\delta _b^c  K^a{}_d -  \delta _d^a K^c{}_b.
\label{(24)}
\end{equation}
The $E_{11}$ algebra also contains an infinite number of
generators of negative level which are partners of those with
positive level but have their indices downstairs;
\begin{equation}
R_{a_1a_2a_3} (-1) , R_{a_1a_2\dots a_6} (-2) , R_{a_1a_2\ldots
a_8,b}(-3), \ldots,
\label{(25)}
\end{equation}
with an identical constraint on the last generator.  The generators of
positive level are associated with negative roots in the Chevalley-Serre
basis. Similarly, those of negative level are associated to positive
roots. Those of zero level contain both positive and negative roots as
well as the entire Cartan subalgebra.
\par
By construction the generators of equations \eqref{(23)} and \eqref{(25)} belong to
representations of GL(11) and  so their commutators with the generators
$K^a{}_b$ are as their index structure suggests. We list the
commutators of the first few generators with $K$ below \cite{w42}:
\begin{gather}
[K^a{}_b, R^{c_1\ldots c_3}]=  3\delta
_b^{[c_1}R^{|a| c_2 c_3]},\label{(26)}\\
[K^a{}_b, R_{c_1\ldots c_3}]= -3 \delta ^a_{[c_1}R_{|b| c_2
c_3]}, \label{(28)}\\
[K^a{}_b, R^{c_1\ldots c_6}]=
6 \delta _b^{[c_1}R^{|a|c_2\ldots c_6]}, \label{(26b)}\\
[K^a{}_b, R_{c_1\ldots c_6}]=  - 6 \delta ^a_{[c_1}R_{|b| c_2\ldots
c_6]}, \label{(28b)}\\
[ K^a{}_b,  R^{c_1\ldots c_8, d} ]=
8 \delta ^{[c_1}_b R^{|a| c_2\ldots c_8], d}  + \delta _b^d
R^{c_1\ldots c_8, a},
\label{(27)} \\
[ K^a{}_b,  R_{c_1\ldots c_8, d} ]=
-8\delta ^a_{[c_1} R_{|b| c_2\ldots c_8], d}  - \delta ^a_d
R_{c_1\ldots c_8, b}.
\label{(29)}
\end{gather}
\par
The commutators of
$E_{11}$ preserve the level, and it turns out that all  the positive
level
generators can just be found from the multiple commutators of $K^a{}_b$ and
$R^{a_1a_2a_3}$ and all the negative  generators from
the multiple commutators of $K^a{}_b$ and
$R_{a_1a_2a_3}$. The commutators of some of the  positive level
generators are given by
\begin{equation}
[ R^{c_1\ldots c_3}, R^{c_4\ldots c_6}]= 2 R^{c_1\ldots c_6},\qquad
[R^{a_1\ldots a_6}, R^{b_1\ldots b_3}]
= 3  R^{a_1\ldots a_6 [b_1 b_2,b_3]}.
\label{(30)}
\end{equation}
Similarly some of the  commutators of the negative definite level
generators are given by
\begin{equation}
[ R_{c_1\ldots c_3}, R_{c_4\ldots c_6}]= 2 R_{c_1\ldots c_6},\qquad
[R_{a_1\ldots a_6}, R_{b_1\ldots b_3}]
= 3  R_{a_1\ldots a_6 [b_1 b_2,b_3]}.
\label{(31)}
\end{equation}
Finally, the commutation relations between the positive and negative
generators of  up  to level three are given by \cite{w31}
\begin{gather}
[ R^{a_1\ldots a_3}, R_{b_1\ldots b_3}]= 18 \delta^{[a_1a_2}_{[b_1b_2}
K^{a_3]}{}_{b_3]}-2\delta^{a_1a_2 a_3}_{b_1b_2 b_3} D, \label{(32)}\\
[ R_{b_1\ldots b_3}, R^{a_1\ldots a_6}]= {5!\over 2}
\delta^{[a_1a_2a_3}_{b_1b_2b_3}R^{a_4a_5a_6]},
\label{(32b)}\\
[ R^{a_1\ldots a_6}, R_{b_1\ldots b_6}]= -5!.3.3
\delta^{[a_1\ldots a_5}_{[b_1\ldots b_5}
K^{a_6]}{}_{b_6]}+5!\delta^{a_1\ldots  a_6}_{b_1\ldots  b_6} D ,
\label{(33)}\\
[ R_{a_1\ldots a_3}, R^{b_1\ldots b_8,c}]= 8.7.2
( \delta_{a_1a_2 a_3}^{[b_1b_2b_3} R^{b_4\ldots b_8] c}-
\delta_{a_1a_2 a_3}^{[b_1b_2 |c|} R^{b_3\ldots b_8]} ),
\label{(34)}\\
[ R_{a_1\ldots a_6}, R^{b_1\ldots b_8,c}]= {7! .2\over 3}
( \delta_{a_1\ldots  a_6}^{[b_1\dots b_6} R^{b_7 b_8] c}-
\delta_{a_1\ldots  a_6}^{c[b_1\ldots b_5 } R^{b_6b_7 b_8]}),
\label{(35)}
\end{gather}
where $D=\sum_b K^b{}_b$. There are similar formulae when
higher or lower level generators are involved.
\par
By examining
the above commutators one can see that the level is nothing more that
the
number of times the generator $R^{a_1a_2a_3}$ minus the number of times
the generator $R_{a_1a_2a_3}$ occurs. For the purposes of this paper
this definition will suffice, but a precise description of  the level is as
follows. Each generator is associated with a root of $E_{11},$ which can
be expressed as a sum of simple roots. Each node of the Dynkin
diagram is
associated with a simple root. The level refers to the GL(11)
decomposition which picks out the eleventh node in figure \ref{fig2}.
Associated to the eleventh node is the simple root $\alpha_{11}$. The
level is the coefficient of $\alpha_{11}$ when the root associated to
that generator is written as the sum of simple roots.
\par
For the purposes of this  paper all that is required to know  about the
$E_{11}$ algebra is the above commutation relations.  The reader who is
interested in  a more detailed account of $E_{11}$ from the
definition of
a Kac-Moody algebra may consult \cite{w42} and the later papers on
$E_{11}$ referenced in this paper. As we will see shortly, the
non-linear realisation of the $E_{11}$ algebra leads to the fields found in the massless bosonic sector of M-theory.
\par
In early papers on $E_{11}$, in addition to  the group element
belonging to $E_{11}$, spacetime was introduced into  the
group element by including a factor of $e^{x^aP_a} $,  where $P_a$
are the
generators of spacetime translations. The generators $P_a$  were
taken to have
non-trivial commutators with the GL(11) generators $K^{a}{}_{b}$ of $E_{11}$, but
trivial commutators with all the non-zero level generators. It was realised from
the beginning that this was an ad hoc and incomplete step.
\par
Later, it was proposed to incorporate spacetime by using  a
representation of $E_{11}$ \cite{w31}, which was denoted by the $l_1$
representation. This representation generalises the notion of spacetime
translation generators.  The
$l_1$ representation, when decomposed into representations of SL(11),
has
the content \cite{w31, w29, w25}
\begin{gather}
L_{A}=\{P_a, (0); Z^{ a_1 a_2} (1) ; Z^{ a_1
\ldots  a_5} (2) ; Z^{ a_1\ldots   a_7,b}(3),
Z^{   a_1\ldots   a_8}(3); Z^{  a_1\ldots   a_8,  b_1 b_2 b_3}(4), Z^{
a_1\ldots    a_9,( b c)}(4),\notag \\ Z^{  a_1\ldots  a_9, b_1 b_2}(4),
   Z^{  a_1\ldots  a_{10}, b}(4), Z^{  a_1\ldots  a_
{11}}(4); Z^{  a_1\ldots  a_9,   b_1\ldots  b_4}(5),
Z^{  a_1\ldots  a_8, b_1\ldots  b_6}(5), Z^{  a_1\ldots
   a_9, b_1\ldots  b_5}(5),\dots  \}.
\label{(36)}
\end{gather}
The numbers in brackets are the levels of the generators which just
counts the number of times the generator $R^{abc}$ acts on the highest
value component in $P_a$. One sees that at the very lowest level it
contains  the spacetime translations
$P_a$ and  then some generators that have the index structure to be
the central charges in the eleven dimensional supersymmetry algebra as
well as an infinite number of higher level objects. From the
mathematical viewpoint, the
$l_1$  representation has the highest weight $\Lambda_1$ which  obeys
the
relations $(\Lambda_1, \alpha_a)=\delta _{a1}$ where $\alpha_a$ are the
simple roots of $E_{11}$. This is just the fundamental representation
associated with node one.  The deduction of the above content,
\eqref{(36)}, from this definition is explained in
\cite{w31, w29, w25}.
\par
At the lowest levels the $l_1$ representation contains objects
that have the correct index structure to be the brane charges; that is
$P_a$,$Z^{ab}$,$Z^{a_1\ldots a_5}\ldots$ associated with the point
particle, M2 brane and M5 brane, respectively. At level three $Z^{ a_1
\ldots   a_7,b}$ probably represents the KK monopole (or D6-brane)
charge. It has been  conjectured
that the $l_1$ representation  contains all  brane charges and there
is now a substantial amount of evidence for this conjecture
\cite{w31, w29, w25, w9}.
\par
The generators of equations \eqref{(23)} and \eqref{(25)}  correspond to
the SL(11) decomposition of $E_{11}$, which is the one appropriate  to
the eleven dimensional theory. To find the theory in $d$ dimensions we
should carry out the  decomposition of the adjoint representation
of $E_{11}$  into representations of the direct product  of the duality
group in $d$ dimensions and GL$(D),$ where $D=11-d$ \cite{w25, w13, w11, w8, w7}. This can be found
from equations \eqref{(23)} and \eqref{(25)} by simply carrying out the
dimensional reduction by hand as will be done in this paper.  Deleting
the $D-$th node, for $D=1, \dots 8,$ we obtain direct products of the
duality groups $E_{10}, E_{9}, E_{8}, E_{7}, E_{6},$ SO(5,5), SL(5) and
SL(2)$\times$SL(3) with GL$(D),$ respectively. The same decomposition is
required for the
$l_1$ representation and the results \cite{w25, w9, w23} are
given in table
\ref{table2}.  Some of the entries in the table agree with those
previously found by taking an explicit charge and using U-duality to
find the other members of the multiplet \cite{EGKR, OPR, OP}.

\begin{table}[htbp]
$$\halign{\centerline{#} \cr
\vbox{\offinterlineskip
\halign{\strut \vrule \quad \hfil # \hfil\quad &\vrule  \quad \hfil #
\hfil\quad &\vrule \hfil # \hfil
&\vrule \hfil # \hfil  &\vrule \hfil # \hfil &\vrule \hfil # \hfil &
\vrule \hfil # \hfil &\vrule \hfil # \hfil &\vrule \hfil # \hfil &
\vrule \hfil # \hfil &\vrule#
\cr
\noalign{\hrule}
D&G&$Z$&$Z^{a}$&$Z^{a_1a_2}$&$Z^{a_1\ldots a_{3}}$&$Z^{a_1\ldots a_
{4}}$&$Z^{a_1\ldots a_{5}}$&$Z^{a_1\ldots a_6}$&$Z^{a_1\ldots a_7}$&\cr
\noalign{\hrule}
8&$SL(3)\otimes SL(2)$&$\bf (3,2)$&$\bf (\overline 3,1)$&$\bf (1,2)$&$\bf
(3,1)$&$\bf (\overline 3,2)$&$\bf (1,3)$&$\bf (3,2)$&$\bf (6,1)$&\cr
&&&&&&&$\bf (8,1)$&$\bf (6,2)$&$\bf (18,1)$&\cr  &&&&&&&$\bf (1,1)$&&$
\bf
(3,1)$&\cr  &&&&&&&&&$\bf (6,1)$&\cr
&&&&&&&&&$\bf (3,3)$&\cr
\noalign{\hrule}
7&$SL(5)$&$\bf 10$&$\bf\overline 5$&$\bf 5$&$\bf \overline {10}$&$\bf 24$&$\bf
40$&$\bf 70$&-&\cr  &&&&&&$\bf 1$&$\bf 15$&$\bf 50$&-&\cr
&&&&&&&$\bf 10$&$\bf 45$&-&\cr
&&&&&&&&$\bf 5$&-&\cr
\noalign{\hrule}
6&$SO(5,5)$&$\bf \overline {16}$&$\bf 10$&$\bf 16$&$\bf 45$&$\bf \overline
{144}$&$\bf 320$&-&-&\cr &&&&&$\bf 1$&$\bf 16$&$\bf 126$&-&-&\cr
&&&&&&&$\bf 120$&-&-&\cr
\noalign{\hrule}
5&$E_6$&$\bf\overline { 27}$&$\bf 27$&$\bf 78$&$\bf \overline {351}$&$\bf
1728$&-&-&-&\cr  &&&&$\bf 1$&$\bf \overline {27}$&$\bf 351$&-&-&-&\cr
&&&&&&$\bf 27$&-&-&-&\cr
\noalign{\hrule}
4&$E_7$&$\bf 56$&$\bf 133$&$\bf 912$&$\bf 8645$&-&-&-&-&\cr
&&&$\bf 1$&$\bf 56$&$\bf 1539$&-&-&-&-&\cr
&&&&&$\bf 133$&-&-&-&-&\cr
&&&&&$\bf 1$&-&-&-&-&\cr
\noalign{\hrule}
3&$E_8$&$\bf 248$&$\bf 3875$&$\bf 147250$&-&-&-&-&-&\cr
&&$\bf1$&$\bf248$&$\bf 30380$&-&-&-&-&-&\cr
&&&$\bf 1$&$\bf 3875$&-&-&-&-&-&\cr
&&&&$\bf 248$&-&-&-&-&-&\cr
&&&&$\bf 1$&-&-&-&-&-&\cr
\noalign{\hrule}
}}\cr
}$$
\caption{ Table giving the representations of the symmetry group $G$ of the form charges in the $l$ multiplet up to and including rank $D-1$ in $D= 8$ dimensions and below \cite{w25, w9, w23}. }
\label{table2}
\end{table}

It was proposed \cite{w31} that the dynamics should
be a non-linear  realisation of semi-direct product of $E_{11}$ and
generators that belonged to the
$l_1$ representation, the motion group of $E_{11}$; denoted by $E_{11}
\ltimes l_1$. This  algebra
contains  the generators of equations \eqref{(23)}, \eqref{(25)} and
those of equation
\eqref{(36)} which  we now take  to be generators and call the
generalised translation generators.
The commutators for the low level generators of the
$l_1$ representation with $R^{a_1a_2a_3}$ are determined up to constants
by demanding that the levels match and so we can take \cite{w31}
\begin{gather}
[R^{a_1a_2a_3}, P_b]= 3 \delta^{[a_1}_b Z^{a_2a_3]}, \label{(37a)} \\
[R^{a_1a_2a_3}, Z^{b_1b_2} ]= Z^{a_1a_2a_3 b_1b_2}, \label{(37b)}  \\
[R^{a_1a_2a_3}, Z^{b_1\ldots b_5} ]=Z^{b_1\ldots b_5[a_1a_2,a_3]}+
Z^{b_1\ldots b_5 a_1a_2 a_3} \label{(37c)}
\end{gather}
The normalisation of the generators is fixed by these relations, see
appendix \ref{A} for a detailed explanation. The
commutators of the generalised translation generators with those of GL(11) are given by
\begin{gather}
[K^a{}_b, P_c]= -\delta _c^a P_b +{1\over
2}\delta _b^a P_c, \label{(38a)}\\
[K^a{}_b, Z^{c_1c_2} ]= 2\delta_b^{[c_1} Z^{|a|c_2]}
+{1\over 2}\delta _b^a Z^{c_1c_2},  \label{(38b)}\\
[K^a{}_b, Z^{c_1\ldots c_5} ]= 5\delta_b^{[c_1} Z^{|a|c_2\ldots c_5]}
+{1\over 2}\delta _b^a Z^{c_1\ldots c_5}.
\label{(38c)}
\end{gather}
Some of the remaining commutators are given by \cite{w31}
\begin{gather}
[R_{a_1a_2a_3}, P_b ]= 0, \label{(39c)}\\
[R_{a_1a_2a_3}, Z^{b_1b_2} ]= 6\delta^{b_1b_2 }_{[a_1a_2} P_{a_3 ]},
\label{(39d)}\\
[R_{a_1a_2a_3}, Z^{b_1\ldots b_5} ]= {5!\over 2} \delta^{[ b_1b_2b_3
}_{a_1a_2a_3} Z^{b_4b_5]}, \label{(39e)} \\
[R_{a_1a_2a_3}, Z^{b_1\ldots b_7, d} ]= {378} \delta^{d[ b_1b_2
}_{a_1a_2a_3} Z^{b_3 \dots b_7]}, \label{(39f)} \\
[R^{a_1\dots a_6}, P_b]= -3 \delta^{[a_1}_b Z^{\ldots a_6]}, \label
{(39a)} \\
[R^{a_1\dots a_6}, Z^{b_1b_2} ]= -Z^{b_1b_2[a_1\ldots
a_5,a_6]}-Z^{b_1b_2 a_1\ldots a_6}. \label{(39b)}
\end{gather}
These commutators can be largely determined by demanding that the level
is preserved and that the Jacobi identities hold. The factor of ${1
\over 2}$ in the terms proportional to $\delta^{a}_{b}$ in equations
\eqref{(38a)}--\eqref{(38c)} are
fixed by the Jacobi identities once it is found to be present in the
first equation, \eqref{(38a)}. These terms follow from the fact that
the
$l_1$ representation is a highest weight representation of $E_{11}$. If
one considers the analogous representation for  subalgebras such as
$E_{10}$, or even the finite dimensional
$E_{n}$ series one finds factors other than ${1\over 2}$. Indeed the
corresponding  factor for $E_n$ is ${1\over
n-9} \; (n \neq 9).$ $E_{9}$ is an exception because it's an affine algebra, so its Cartan matrix has vanishing determinant.
\par
To carry out explicit computations of the $E_{11}\ltimes l_1$
non-linear realisation at
low levels, one only needs the above commutators
and one does not have to absorb the general theory of Kac-Moody
algebras.
\par
As explained in the introduction the non-linear realisation we are using
here is not a sigma model as the $l_1$ representation are generators
associated with spacetime and they introduce the coordinates of the
generalised spacetime into the theory. How to construct such non-linear
realisations is illustrated by example in \cite{w45, w42} and many
of the later papers on $E_{11}$ even though only the generators of
spacetime translations $P_a$ are used. The non-linear realisation of
$E_{11}\ltimes l_1$ was used in \cite{w11} to construct all five
dimensional gauged supergravities and in \cite{w3} and \cite{w1} to
construct the IIA ten dimensional supergravity in the NS-NS and R-R
sectors respectively. The next section uses it to construct the four
dimensional theory at level zero but keeping only the scalar coordinates
in the ten of SL(5). That section may be read at the same time as the abstract material below. The material in this section may also be compared with section \ref{2} which covers the SL(5) case without the complications of the $E_{11} \ltimes l_1$ algebra.
\par
The non-linear realisation is built from the group
element
\begin{equation}
g=g_lg_E.
\label{(40)}
\end{equation}
In eleven dimensions the group element $g_E$ takes the form
\begin{equation}
g_E= \ldots  e^{{1\over6!} C^{a_1 \dots a_6}R_{a_1 \dots a_6}}e^{{1
\over 3!} C^{a_1a_2a_3}R_{a_1a_2a_3}}e^{h_a{}^b K^a{}_b}
e^{{1\over 3!} C_{a_1a_2a_3}R^{a_1a_2a_3}}e^{{1\over6!} C_{a_1 \dots
a_6}R^{a_1 \dots a_6}}\ldots.
\label{(41)}
\end{equation}
Using equation \eqref{(36)}, the group element $g_l$,  in eleven
dimensions, takes the form
\begin{equation}
g_l=e^{x^a P_a} e^{{1 \over \sqrt 2} x_{ab}Z^{ab}}e^{{1\over \sqrt 5}x_
{a_1\ldots
a_5}Z^{a_1\ldots a_5}}\ldots.
\label{(42)}
\end{equation}
The precise choice of the normalisation is explained in appendix A.
\par
Thus the non-linear realisation  of $E_{11}\ltimes l_1$ introduces a
generalised spacetime with coordinates \cite{w31, w23, w11}
\begin{gather}
z^{\Pi}=\{ x^a, ; x_{ a_1 a_2}; x_{ a_1
\ldots  a_5}; x_{ a_1\ldots   a_7,b},
x_{   a_1\ldots   a_8}; x_{  a_1\ldots   a_8,  b_1 b_2 b_3}, x_{
a_1\ldots    a_9,( b c)}, x_{  a_1\ldots  a_9, b_1 b_2},   \notag \\
   x_{  a_1\ldots  a_{10}, b}, x_{  a_1\ldots  a_
{11}}; x_{  a_1\ldots  a_9,   b_1\ldots  b_4, c},
x_{  a_1\ldots  a_8, b_1\ldots  b_6}, x_{  a_1\ldots
   a_9, b_1\ldots  b_5},\dots  \},
\label{(43)}
\end{gather}
where the first coordinate $x^a$ is the coordinate of the
spacetime we are so used to. However the multiplet contains an
infinite number of additional coordinates.  As a result of the way they
have arisen, there is a one to one correspondence between the generators
of equation
\eqref{(36)} and the coordinates of equation \eqref{(43)}  so that each
coordinate is automatically associated with a brane charge. In
particular,  the usual coordinates
$x^a$ are associated with the generators $P_a$ of spacetime
translations, the   coordinates $x_{ab}$ with the charge $Z^{ab}$
of the M2 brane and so on. One can show \cite{w29} that  for every field
there is a corresponding brane charge, for example
$h_a{}^b, C_{a_1 a_2 a_3},\ldots $ correspond to
$P_a, Z^{ab},\ldots $,  respectively.   As a result every field now
has a corresponding coordinate associated with it;
we can think of the usual spacetime coordinates $x^a$ as being
associated with the metric, the coordinates $x_{ab}$ as associated with
the three form field $C_{a_1a_2a_3}$, etc. Thus this construction
generalises spacetime to take account of the objects within it.
Einstein's theory corresponds to the lowest level.
We take the fields
$h_a{}^b, C_{a_1a_2a_3},\ldots $ to depend on all of the coordinates
$x^a,$ $x_{ab}$ etc. Introducing the generator $P_a$ on its own, as
mentioned
above, is  just the lowest order approximation.
\par
The group element in lower dimensions is easily written down using the
generators of $E_{11}$ as decomposed into representations of
$GL(D)\otimes E_{d} $ where $D=11-d$. As mentioned above, elements of
$l_1$ are given in table \ref{table2}. We find, in table \ref
{table2}, the   scalar, vector, and
higher rank  generators  in $D-$dimensions contained in the
$l_1$ representation.  In particular, we find that the
scalar charges in the $l_1$ representation  in $d=4,5,6,7$
dimensions belong to the 10, $\overline {16}$, $\overline {27}$ and 56 representations of
SL(5), SO(5,5), $E_6$ and $E_7$ respectively \cite{w25, w23}.
In this paper we will be interested in the  non-linear
realisation  at level zero with respect to the deleted node. With  this
restriction only a finite number of fields and coordinates will remain.
\par
A non-linear realisation is specified by a choice of algebra  and
subalgebra, called the local subalgebra. In our case the algebra is
$E_{11}$ and we will denote the local subalgebra by $I(E_{11})$.  By
definition, the non-linear realisation is just a dynamics which is
invariant under the transformations
\begin{equation}
g\to g_0 g,\  g_0\in E_{11}\ltimes l_1,\quad {\rm and } \quad g\to gh,\
h\in I(E_{11})
\label{(44)}
\end{equation}
In this equation $g_0$ is a rigid transformation, and so does not depend
on the generalised spacetime,  while $h$ is a local transformation
which
does depend on the generalised
spacetime. The local subalgebra
$I(E_{11})$ is taken to be a maximal subalgebra that is invariant
under Cartan involution. This subalgebra of $E_{11}$ is generated by
\begin{equation}
K^a{}_b -\eta_{bc}\eta ^{ad}K^c{}_d, \quad R^{a_1a_2a_3}
- \eta^{a_1b_1}\eta^{a_2b_2}\eta^{a_3b_3} R_{b_1b_2b_3}, \quad R^
{a_1a_2\dots
a_6} +\eta^{a_1b_1}\ldots \eta^{a_6b_6} R_{b_1 b_2\dots
b_6}  ,\ldots,
\label{(45)}
\end{equation}
where $\eta$ is the Minkowski metric.
The Cartan involution invariant subgroups of the groups SL$(n)$, SO$(n,n)$,
$E_6$ and $E_7$ are their maximally compact subgroups, which
are SO$(n),$ $SO(n)\otimes
SO(n)$, USp(8) and SU(8) respectively, provided the $d$ dimensions are all spacelike. Hence
at the lowest level the local subalgebra is just the Lorentz group.  We
may therefore use the local transformation of equation \eqref{(44)}
to  bring
the group element
$g_E$ in eleven dimensions into the form
\begin{equation}
g_E= e^{h_a{}^b K^a{}_b} e^{{1\over 3!} C_{a_1a_2a_3} R^{a_1a_2a_3}}
\ldots.
\label{(46)}
\end{equation}
This mostly contains the generators of  the Borel subalgebra of
$E_{11}$ which are the generators given in equation \eqref{(23)}. The
exception
is the field at level zero, i.e. $h_a{}^b$ where we have chosen not
to fix all of the local Lorentz group. The Cartan involution $I$
takes, up to a sign,  a  generator with a positive level to a  generator
with a negative level and with the same set of indices but downstairs,
that is it takes a contravariant to a contragredient
SL(11) representation. More technically it takes a generator with a
positive root
$\alpha$  to a generator with the negative root $-\alpha,$ for
example $I(R^{a_1 \dots a_3})=-R_{a_1 \dots a_3} $ and $I(R^{a_1
\dots a_6})=R_{a_1 \dots a_6}.$
Furthermore, it maps the generators of the Cartan subalgebra into
themselves. For
a more formal definition see,  for example,  \cite{w42} and many later
papers on $E_{11}$ .
\par
As we explained to find the non-linear realisation in  eleven dimensions
we delete node eleven and decompose $E_{11}\ltimes l_1$. At level zero
this  algebra becomes
$GL(11)\ltimes P_\mu$ where $P_\mu$ are  just the usual spacetime
translations in eleven dimensions. At level zero  $I(E_{11})$
is just the Lorentz group. Thus in this case the generalised spacetime
has the   coordinates $x^a$ and so is just our familiar spacetime. The
only  fields are $h_a{}^b$.   In fact the
non-linear realisation, after the adjustment of a few constants that are
not determined,  leads to eleven dimensional gravity. It turns out that
$e^h$ viewed as a matrix is just the vielbein \cite{BO} just as was shown
in section \ref{2}. In what follows it will
be useful to recall that the non-linear realisation of the semi-direct product of
$GL(d)$ and spacetime
translations leads to
$d$-dimensional gravity,  as was shown long ago for the case of four
dimensions \cite{BO}.
\par
Under a  rigid $g_0\in E_{11}$ and a local $H \in I(E_{11}),$  the
different parts of the group element transform as
\begin{gather}
g_l\to g_0 g_l
(g_0)^{-1}, \quad {\rm and} \quad
g_E\to g_0g_E \label{(47)}\\
g_l\to g_l, \quad {\rm and} \quad
g_E\to g_E h,
\label{(48)}
\end{gather}
respectively, as the $l_1$ generators form a realisation of
$E_{11}$.  As a result the coordinates transform under $G$ as
\begin{equation}
z^\Pi L_\Pi\to g_0 z^\Pi L_\Pi (g_0)^{-1}
\label{(49)}
\end{equation}
\par
To give a more concrete meaning to the above rigid transformations we
will
carry them out for $g_0=e^{{1\over3!} a_{a_1a_2a_3}R^{a_1a_2a_3}}$ where
$a_{a_1a_2a_3}$ is a constant parameter. Using equation \eqref{(47)} and
equations \eqref{(42)} and \eqref{(46)}, we find that
\begin{gather}
\delta x^a=0, \qquad  \delta x_{ab}= {1\over{\sqrt{2}}} a_{abc} x^{c} ,
\qquad  \delta h_a{}^b= 0 , \notag \\
\delta
C_{a_1a_2a_3}= a_{a_1a_2a_3} - 3 a_{b[a_1a_2} h_{a_3]}{}^{b} ,\qquad
\delta C_{a_1\dots a_6}=0.
\label{(50)}
\end{gather}
\par
To construct the dynamics from
the  non-linear realisation,  it is usual to first construct the Cartan
form. The  Cartan form belongs to the Lie algebra and so  in our case
the algebra
$E_{11}\ltimes l_1$. As such,  it  can be written as
\begin{equation}
{\cal V}\equiv
g^{-1}d g= dz^\Pi E_\Pi{}^A L_A+ dz^\Pi G_{\Pi,*} R^*
\label{(51)}
\end{equation}
where $L_A$ are the generators of the $l_1$ representation and  $R^*$ are the
generators of $E_{11}$ in equation \eqref{(23)} and \eqref{(46)} with $*$ denoting  the
appropriate set of indices. When we write the sums involving the $L_A
$ generators
we are including  the square root of the combinatorial factors that
occur in the group element in equation \eqref{(42)}.
Since the generators $L_A$ form a representation of
$E_{11},$ the Cartan form is given by
\begin{equation}
{\cal V}= g^{-1}_E dz^A L_A g_E+ g^{-1}_Ed g_E
\label{(52)}
\end{equation}
where we have  assumed that the generators $L_A$ mutually commute. We
may write
\begin{equation}
dz^\Pi E_\Pi{}^A L_A= g^{-1}_E dz^A L_A g_E=dz^T\cdot  E\cdot  L
\label{(53)}
\end{equation}
where in the last line we have used an obvious matrix notation in that
the matrix $E$ has the elements $E_\Pi {}^A$.  The remaining part of the
Cartan form is given by
\begin{equation}
dz^\Pi G_{\Pi,*} R^*= g^{-1}_E d g_E
\label{(54)}
\end{equation}
and it is just the Cartan form of $E_{11}$.
\par
The Cartan form \eqref{(51)} is  obviously inert under the rigid $g_0
$ transformations
of equation \eqref{(44)}.
Note that the generators of the $l_1$
representation can carry either a
$\Pi$ or an $A$ index depending on the context; this is not a change carried out
with the vielbein and
$L_A=L_\Pi\delta^\Pi_A$.
As the $l_1$ generators  form a representation of $E_{11}$ it
follows that
$dz^\Pi E_\Pi{}^A$ and $dz^\Pi G_{\Pi, *}$ are separately invariant
under these rigid transformations.  However, the coordinates,  and so
$dz^\Pi$,  do transform under $g_0$ and as a result  $E_\Pi{}^A$  and
$G_{\Pi, *}$  are not invariant under $g_0$ transformations.
\par
Under a local  transformation
$g\to gh$  of equation \eqref{(51)} the Cartan forms transform  as $
{\cal
V}\to h^{-1} {\cal V}h+h^{-1}d h$.
To find quantities that only transform under the local
subalgebra we can rewrite ${\cal V}$ as
\begin{equation}
{\cal V}= g^{-1}d g= dz^\Pi E_\Pi{}^A( L_A+  G_{A,*} R^*)
\label{(55)}
\end{equation}
where we recognise that $G_{A,*}=(E^{-1})_A{}^\Pi  G_{\Pi,*}$.
Since  $dz^\Pi E_\Pi{}^A$ and $R^{\star}$ are inert under rigid $g_0$
transformations, it follows that
$G_{A,*}$ are also inert under
$g_0$ transformations and just transform under local transformations. As
such they  are useful quantities with which to construct the dynamics as
one need only solve the problem of finding  objects which are
invariant under the local symmetry. For objects in the coset
directions of ${\cal V}$,  the $h^{-1}d h$ terms in the
transformation of
equation \eqref{(51)} are absent and we   may think of
$G_{A,*}$  as  transforming covariantly---in effect they are the
covariant derivatives of the fields. Thus  working with the Cartan forms
one only has to solve the problem of find invariants   under the local
transformations $h$.  In fact the situation is a little more subtle as
we have used the local subgroup to choose our group element to belong
to the Borel subgroup and  then a
$g_0$ transformation requires a local compensating transformation.
However, as the final dynamics  is invariant under local
transformations these are  automatically taken care of.
\par
Since the generators of the $l_1$ representation transform as a
representation of $E_{11}$ we can write equation \eqref{(49)} as
\begin{equation}
z^\Pi L_\Pi\to g_0 z^\Pi L_\Pi (g_0)^{-1}= z^\Pi
D(g_0^{-1})_\Pi{}^\Lambda  L_\Lambda
\label{(56)}
\end{equation}
where $D(g_0^{-1})_\Pi{}^\Lambda $  is the corresponding matrix
representation. More formally we can  define the action of the  $l_1$
representation  of
$E_{11}$, to which the generators $L_\Pi$  belong,    by
$$U(k)( L_\Pi  )\equiv k^{-1} L_\Pi k= D(k)_\Pi{}^\Lambda L_\Lambda, $$
where
$k\in E_{11}$. As a result we find that in matrix notation
$$dz^T \to dz^{T\prime}= dz ^T D(g_0^{-1}),$$ or putting in the indices
\begin{equation}
dz^{ \Pi}
\to dz^{ \Pi\prime }= dz^{ \Lambda}D(g_0^{-1})_\Lambda{}^\Pi.
\label{(57)}
\end{equation}
Consequently, the derivative
$\partial_\Pi= {\partial\over \partial z^\Pi}$ in the generalised
spacetime transforms as $$\partial_\Pi^\prime=
D(g_0)_\Pi{}^\Lambda\partial_\Lambda.$$ Examining equation \eqref{(53)}, we
note that the generalised vielbein $E$ in matrix form is given by
\begin{equation}
E_\Pi{}^\Lambda= D(g_E)_\Pi{}^\Lambda.
\label{(58)}
\end{equation}
\par
As the Cartan form is inert under rigid transformations,  its  action on
the coordinates must be  compensated by a corresponding change on the
lower index of $E$, using equation \eqref{(57)},  we find this to be
given by
$E_\Pi{}^{A\prime}= D(g_0) _\Pi{}^\Lambda E_\Lambda{}^{A}$. Thus the
lower index is a world index, while
$E_\Pi{}^{A\prime} $ transforms on its upper index by a local
$h$  transformation and so we can  think of the upper
index as a tangent index.
Consequently, we  can think of  $E_\Pi {}^A$ as a generalised
vielbein which controls the geometry of the generalised spacetime.
\par
In almost all the $E_{11}$ papers the dynamics has been constructed
using
the Cartan forms. However, one can also proceed in another way and this
was done in \cite{davidmalcolm, w3} which we now follow.  Let us define
\begin{equation}
M\equiv g_E I_c(g_E^{-1}),
\label{(59)}
\end{equation}
   where $I_c$ is the Cartan involution.
It is easy to see, using equation \eqref{(48)} that $M$ is inert
under local
transformations as by definition $I_c (h)=h$. However, under a rigid
transformation  $M\to M^\prime = g_0 M I_c(g_0^{-1})$
under rigid transformations.  Using $E=D(g_E),$ we find that $M$ in
the
$l_1$ representation is given by
\begin{equation}
D(M)=D(g_E)D(I_c(g_E^{-1}))=E E^{\#}
\label{(60)}
\end{equation}
where
$E^{\#}= I_c(D(g_E^{-1}))$, which for many
groups it is just the transpose. Writing out the indices
explicitly we find that
\begin{equation} D(M)_{\Pi
\Lambda}=(EE^{\#})_{\Pi
\Lambda}
\label{(61)}
\end{equation} and we can write its rigid transformation as
\begin{equation} D(M)_{\Pi
\Lambda}\to   D(g_0^{-1})_\Lambda{}^\Gamma D(M)_{\Gamma
\Theta}D(I_c(g_0^{-1}))_\Pi {}^\Theta.
\label{(62)}
\end{equation}
Using this method, the problem of finding invariants reduces to
constructing $g_0$ invariants from $M$. In the subsequent sections we
will carry this procedure out in detail for the various dimensions.  If
we restrict ourselves to two spacetime derivatives then the most
general
invariant  Lagrangian up to boundary terms is given by \cite{davidmalcolm, w3}
\begin{gather}
L=c_1 M^{ST} \partial_S M^{PQ} \partial_T M_{PQ}
+c_2 M^{ST} \partial_S M^{PQ} \partial_P M_{TQ} +c_3 M^{MN} M^{ST}(M^{PQ} \partial_S M_{PQ}) (\partial_M M_{NT}) 
\notag \\
+c_4  M^{ST} (M^{MN} \partial_S M_{MN}) (M^{PQ} \partial_T M_{PQ}) +c_5 M_{RQ} \partial_{S} M^{SR} \partial_{P} M^{PQ} 
\label{(63)}
\end{gather}
where $c_1,\ldots ,c_5$ are constants, and $M^{ST}$ denotes $(M^{-1})^{ST}$. The term with coefficient $c_5$ never gives rise to a U(1) gauge-invariant result. One can therefore set $c_5$ equal to zero with impunity. Boundary terms may be included in terms of the generalised metric \cite{bermanmusaevperry}.
\par
The non-linear realisation  introduces the
generalised spacetime,  but since it also specifies the dynamics,  at
least up to a few constants,  it also determines the  geometry of the
generalised spacetime. However, it is important to understand that the
non-linear realisation as described above,  and used in the papers on
$E_{11}$,  is not what is usually described as a sigma model. The latter
corresponds to a non-linear realisation in which the group contains no
generators associated with any spacetime. As a result the
coordinates are introduced by hand and act as dummy variables upon which
the fields depend. In contrast the non-linear realisation described
here
has generators which lead to the introduction of spacetime into the
group
element  and so the generalised spacetime plays  a central role in the
way the dynamics is formulated.
\par
We note that the conjectured theory based on $E_{10}$ \cite{e10} is
quite different. It uses a
non-linear realisation that is equivalent to that which  is usually
known
as a sigma model. In this formulation the fields only depend on time and
it is hoped that spacetime will emerge at  higher levels in the
algebra.
\par
The Lagrangian  of equation \eqref{(63)} contains five undetermined
constants
and, since  it is to be integrated over a generalised spacetime, it is
of a rather unfamiliar form. One may like to find a theory that contains
only the spacetime that is familiar to us. Although up to this stage
the procedure has been very systematic, how to proceed further is not
completely clear. One approach used in the non-linear realisation of
$GL(D)\ltimes I^4$ to find gravity \cite{BO} is to demand some extra
symmetries such as conformal symmetry. This step, taken together with
the
original non-linear realisation  is equivalent to demanding general
coordinate invariance. This procedure was also followed in the $E_{11}$
approach \cite{w45, w42} and many subsequent papers. This
procedure has been generalised in the work of \cite{hillmann} which
considered the non-linear realisation of $E_{11} \ltimes l_1$ applied to
seven dimensions. In \cite{hillmann} the field dependence on the
resulting generalised spacetime was restricted to be only over the
usual
coordinates of spacetime and then the action was required to be
invariant under general coordinate invariance and gauge symmetries.
This is the strategy we will adopt here. One finds in all known cases
that one can adjust the constants so that this is possible.

\section{Four  Dimensions: SL(5) revisited}
\label{4}

\begin{figure}[http]
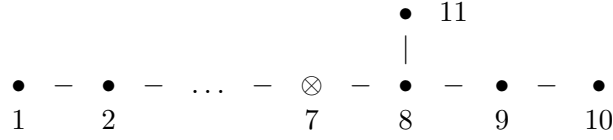

$$
\begin{array}{ccccccccccccccccc}
     & &       & &         & &       &  &\bullet&11&       & &        \\
     & &       & &       & &       &  &   |   &  &          &
&         \\
\bullet &-&\bullet&-&\ldots &-&\otimes&- &\bullet&- &\bullet&-&
\bullet \\
    1    & &   2      & &         & &   7   &  &   8   &  &   9   &
&    10    \\
\end{array}
$$
\caption{The $E_{11}$ Dynkin diagram appropriate to four dimensions}
\label{fig3}
\end{figure}

In this section, we carry out the non-linear realisation of
$E_{11}\ltimes l_1$ appropriate to four dimensions at the lowest
level. That is we will systematically carry out the method given in the
previous section applied to this case. To find  the four-dimensional
theory we delete node seven of the
$E_{11}$ Dynkin diagram to leave the algebra
$GL(7)\otimes SL(5)$,  see figure \ref{fig3}, and decompose $E_{11}
\ltimes l_1$
into this subalgebra. The subalgebra SL(5)
is the well known duality group in the reduction to seven dimensions.

In this paper we are interested in the lowest level
result. The simplest
way to find the low level algebra is to carry out by hand the
dimensional
reduction on the generators of $E_{11}\ltimes l_1$ given in equations
\eqref{(23)}, \eqref{(25)} and \eqref{(36)}. Letting $i,j,\dots
=1,2,3,4$ be the indices
corresponding to the four dimensions we find that the only generators of
$E_{11}$, \eqref{(23)} and \eqref{(25)},  that remain are
\begin{equation}
K^i{}_j, R^{i_1i_2i_3},  R_{i_1i_2i_3}\quad {\rm and } \quad
K^a{}_b, \ a,b=1,2\ldots 7
\label{(64)}
\end{equation}
of GL(7). We are using the convention that $i,j,k,\ldots $ are
tangent indices in
the four dimensional space and  $a,b,c,\ldots $ are tangent indices in
the seven dimensional space.   The generators listed in \eqref{(64)}
have level zero.  We
observe that  the level zero generators have no mixed indices.  For the
decomposition corresponding to deleting node seven, the generators
$K^i{}_a$ ($K^a{}_i$)  have  level
1 (-1) and multiple commutators of these generators together with the
above
generators at level zero will lead to all of the $E_{11}$ Kac-Moody
algebra. More technically a generator has level $n$ if its
corresponding root, when expressed in terms of simple roots,
contains the
simple root $\alpha_7$ with factor $n$.
\par
Keeping only level zero generators we find, using equations \eqref{(24)}, \eqref{(26)}, \eqref{(28)} and
\eqref{(32)}, that the generators of equation \eqref{(64)}  obey the
algebra
\begin{gather*}
[K^i{}_j, K^k{}_l]=\delta_j^k K^i{}_l-\delta_l^i K^k{}_j,\\
[K^i{}_j , R^{k_1k_2k_3}]= 3\delta^{[k_1}_j R^{|i|k_2k_3]} , \\
[K^i{}_j , R_{k_1k_2k_3}]=- 3\delta_{[k_1}^i R_{|j|k_2k_3]} ,
\\
[ R^{i_1i_2i_3}, R_{j_1j_2j_3} ]= 18\delta^{[i_1i_2}_
{[j_1j_2} K^{i_3]}_{j_3 ]}-2\delta^{i_1i_2i_3}_
{j_1j_2j_3} (\sum_j K^j{}_j+\sum_a K^a{}_a)
; \\  [K^a{}_b , K^d{}_c ]=\delta^d_b K^a{}_c -
\delta^a_c K^d{}_b
\end{gather*}
with all remaining commutators being zero. To see that this really is
the
algebra $GL(7)\otimes SL(5)$ we should redefine the generators of SL(4) to
be $\tilde K^i{}_j= K^i{}_j -{1\over 5} \delta^i_j \sum_a K^a{}_a$
and then  the generators $\tilde K^i{}_j, R^{i_1i_2i_3}$ and
$R_{i_1i_2i_3}$ generate SL(5). The generators
$K^a{}_b, \ a,b=1,2\ldots 7$ obey the algebra of  GL(7) and
commute with those of SL(5).
\par
The generators of SL(5) are contained in the
generators $M^I{}_J, I,J=1\ldots , 5$, the identification with those
above being
\begin{equation}
M^I{}_J= \begin{cases} M^i{}_j=\tilde K^i{}_j-{1\over 3} \sum_k \tilde
K^k{}_k,\quad & i,j=1,\ldots ,4 \\
M^i{}_5={1\over 3!}\epsilon^{ij_1j_2j_3} R_{j_1j_2j_3}\quad &
j_1,j_2,j_3=1,\ldots ,4 \\
M^5{}_i={1\over 3!}\epsilon_{ij_1j_2j_3}R^{j_1j_2j_3}\quad &
j_1,j_2,j_3=1,\ldots ,4 \\ \end{cases},
\label{(66)}
\end{equation}
whereupon we find the standard algebra of SL(5), namely
\begin{equation}
[M^I{}_J,M^K{}_L]=\delta^K_J M^I{}_L -\delta^I_L M^K{}_J.
\label{(67)}
\end{equation}
Since the SL(5) generators $M^I{}_J$ are traceless we have defined
$M^5{}_5=-\sum_{i=1} M^i{}_i$.
\par
We now consider the $l_1$ representation at lowest level. Carrying out
the dimensional reduction on equation \eqref{(36)} we find that it
contains
\begin{equation}
P_i, Z^{ij}, \quad  i, j =1,2,3,4\qquad {\rm and }\qquad P_a, \quad
a=1,2\ldots 7.
\label{(68)}
\end{equation}
The commutators of the
generators of equation \eqref{(68)} are found using equations \eqref
{(37a)}, \eqref{(37b)}, \eqref{(38a)}, \eqref{(38b)}, \eqref{(39c)},
\eqref{(39d)}
to be
\begin{gather}
[K^i{}_j , P_l]= - \delta _l^i P_j+{1\over 2}\delta ^i_j P_l, \label
{(69a)}\\
[K^i{}_j , Z^{kl}]= 2\delta ^{[k}_j Z^{|i|l]}+{1\over 2}\delta ^i_j Z^
{kl}, \label{(69b)} \\
   [R^{i_1i_2i_3}, P_j]= 3\delta ^{[i_1}_j Z^{i_2i_3]},\label{(69c)}\\
[R^{i_1i_2i_3}, Z^{kl}]=0 ,\label{(69d)}\\
[R_{i_1i_2i_3}, P_j]=0 , \label{(69e)}\\
[R_{i_1i_2i_3}, Z^{jk}]=6\delta^{jk}_{[i_1i_2}P_{i_3]};\label{(69f)}\\
[K^a{}_b , P_c ]= -\delta_c^a P_b+{1\over 2} \delta_b^a P_c
\label{(69g)}
\end{gather}
as well as
\begin{equation}
[K^a{}_b , P_l]= {1\over 2}\delta ^a_b P_l,\qquad\qquad
[K^a{}_b , Z^{ij}]= {1\over 2}\delta ^a_b  Z^{ij},\qquad\qquad
[K^i{}_j , P_a]= {1\over 2}\delta ^i_j P_a.
\label{(70)}
\end{equation}
All the remaining commutators are zero. We also take all the generators
in the $l_1$ representation to commute with themselves.
\par
We can package the generators of equation \eqref{(68)} with $i,j,
\ldots $
indices into
$P_{IJ}=-P_{JI}, \ I,J=1,\ldots ,5,$ where
\begin{equation}
P_{IJ}= \begin{cases} P_{i5}=P_i&\quad \ i=1,\ldots ,4 \\
P_{ij}= {1\over 2} \epsilon_{ijkl}Z^{kl}&\quad \ i,j,k,l=1,\ldots ,
4 \\ \end{cases}.
\label{(71)}
\end{equation}
Using equations \eqref{(69a)}--\eqref{(69f)}, the  commutator of $P_
{IJ}$ with the generators
of SL(5)  can be written as
\begin{equation}
[M^I{}_J,P_{LM}]=-\delta _L^IP_{JM}-\delta _M^IP_{LJ} +{2\over
5}\delta^I_J P_{LM}.
\label{(72)}
\end{equation}
We recognise that the generators $P_{LM}$ belong to the 10-dimensional
representation of SL(5). Furthermore, one finds that $[M^I{}_J, P_a]
=0,$ hence the $P_{a}$ are SL(5) singlets,  but transform as the 7-dimensional
representation of GL(7). This is very similar to what is done in
section \ref{2}. The difference being that here the algebra is
derived from $E_{11}.$ We find that it includes the extra seven
dimensions of spacetime, and some numerical factors in the algebra
are different. In particular, comparing the equations in \eqref{(8)}
to equations \eqref{(69a)} and \eqref{(69b)}, the coefficient of the
terms proportional to $\delta^{i}_{j}$ are different, $-1/5$ and $1/2
$ respectively.
\par
At level zero the non-linear realisation of
$E_{11}\ltimes l_1$ reduces to the non-linear realisation of
$(GL(7)\otimes SL(5))\ltimes (P_a\oplus
P_{IJ})$. The local subalgebra is
generated by
$K^a{}_b-\eta^{ad}\eta_{bc}K^c{}_d$ and
$K^i{}_j-K^j{}_i$ and  $R^{ijk}-R_{ijk}$ respectively. The use of the
Minkowski metric $\eta_{ab} $ to define the local subalgebra leads to
the subgroup SO(1,6) rather than SO(7). Thus the local
subalgebra is $SO(1,6)\otimes SO(5)$. In fact   SO(7) and SO(5) are the
standard Cartan involution invariant subalgebras of GL(7) and SL(5) and
using the Minkowski metric for the first group results from using  a
slightly different Cartan involution. The non-linear realisation is
built
from the group element
$g_lg_E$  of equation \eqref{(40)} now restricted to level zero.
Taking into
account the local symmetry,  the
$GL(7)\otimes SL(5)$ part of the group element can be written as
\begin{equation}
g_E^{(0)}= e^{h_i{}^j K^i{}_j}e^{{1\over
3!} C_{i_1i_2i_3}R^{i_1i_2i_3}}e^{\hat h_a{}^b K^a{}_b}.
\label{(73)}
\end{equation}
The superscript 0 just indicates we are at level zero. Hence we find
that
the non-linear realisation introduces the fields
\begin{equation}
h_i{}^j, C_{i_1i_2i_3},\quad {\rm and }\quad  \hat h_a{}^b.
\label{(74)}
\end{equation}
We note that the field $C_{i_1i_2i_3}$ was always denoted as
$A_{i_1i_2i_3}$ in the previous literature on $E_{11}$.
\par The part of the group element
arising from the
$l_1$ representation  is given by
\begin{equation}
g_l^{(0)}= e^{x^iP_i + {1\over \sqrt 2} {x_{ij} Z^{ij} }} e^{x^a P_a}.
\label{(75)}
\end{equation}
As such we see that the
$E_{11}\ltimes l_1$ non-linear realisation at level zero
introduces a    generalised
spacetime with the coordinates
\begin{equation}
x^i, x_{ij} \quad {\rm and } \quad  x^a, \quad a=1,\ldots , 7.
\label{(76)}
\end{equation}
The last coordinates are just the usual seven
dimensional spacetime and belong to the $\overline 7$-dimensional
representation of GL(7). The first set of coordinates of
equation \eqref{(76)}  are associated with the spacetime
translation  and
the membrane charges, respectively, and transform as a $\overline {10}$  of
SL(5); we could write them as $X^{IJ}, I,J=1,2,\ldots ,5$. The fields of
equation \eqref{(74)} are taken to depend on the coordinates of equation
\eqref{(76)}.  Thus at lowest level the non-linear realisation
involves the
group element
\begin{equation}
g^{(0)}=g_l^{(0)} g_E^{(0)}=e^{x^iP_i+{1\over \sqrt 2} {x_{ij} Z^{ij} }}
e^{x^a P_a} e^{h_i{}^j K^i{}_j}e^{{1\over 3!}
C_{i_1i_2i_3}R^{i_1i_2i_3}}e^{\hat h_a{}^b K^a{}_b}.
\label{(77)}
\end{equation}
\par
If would be interesting to construct this non-linear realisation; one
would find gravity in seven dimensions coupled to a part that is
the non-linear realisation of $SL(5)\ltimes P_{IJ}$.
However, in
this paper we will  consider a  simplified   non-linear
realisation. In a future paper, we will discuss how the other
components of the metric and $C$ appear in the non-linear realisation
and the action. We note that the generators of SL(5) commute with
those of
GL(7) and the seven dimensional spacetime translations, i.e with $IGL
(7)=
GL(7)\ltimes \{P_a\}$. Indeed the only non-trivial commutator between
$SL(5)\ltimes\{P_i, Z^{ij}\}$ and IGL(7)   is that of the
generators of GL(7) which scale the $\{P_i, Z^{ij}\}$ generators by a
${1\over 2}$ factor. As such the  $SL(5)\ltimes \{P_i,
Z^{ij}\}$ transformations of the non-linear realisation do not affect
the
parts of the group element belonging to IGL(7), that is they do not
affect the spacetime coordinate $x^a$ of the gravity field
$\hat h_a{}^b$. As such it is consistent to set the IGL(7) part of the
non-linear realisation to zero, that is set $x^a=0=\hat h_a{}^b$.
This means that we can just
consider the non-linear realisation  of
$SL(5))\ltimes \{P_i, Z^{ij}\}$ whose corresponding group element is
given by
\begin{equation}
g^{(0)\prime}=e^{x^iP_i+ {1\over \sqrt 2} {x_{ij} Z^{ij} } }  e^{h_i{}
^j K^i{}_j}e^{{1\over
3!} C_{i_1i_2i_3}R^{i_1i_2i_3}}=
g_l^{(0)\prime} g_E^{(0)\prime}
\label{(78)}
\end{equation}
The prime corresponds to the fact that we have dropped the generators
$P_a, K^a{}_b$ and the coordinate $x^a$ and field
$\hat h_a{}^b$.  The remaining fields, namely $h_i{}^j$ and $
C_{i_1i_2i_3}$  now   only
depend on the coordinates $x^i$ and $x_{ij}$.  We note that this
would not
be possible if one were to consider
$E_{11}\ltimes l_1$ at higher levels, nonetheless the results provide an
interesting laboratory in which to study the generalised spacetime
introduced in the non-linear realisation.
\par
Usually when carrying out a Kaluza-Klein reduction to seven dimensions one  neglects  the
dependence of the fields on the spacetime coordinates associated
with the
upper four dimensions leaving the fields to depend on the seven
dimensional spacetime. However, as discussed previously in this paper,
a   different approach was adopted  in the papers \cite{davidmalcolm, BGP} where one
neglected the dependence on the seven dimensions and kept a
dependence on
the coordinates associated with the upper space. The simplification of
the non-linear realisation we have just carried out corresponds to this
latter approach.
\par
It is now straightforward to construct the non-linear realisation. The
vielbein on the generalised spacetime is given by equation \eqref{(53)}
which in this case becomes
\begin{equation}
dz\cdot E\cdot L= (g_E^{(0)\prime})^{-1}(dx^iP_i+ {1\over \sqrt
2}dx_{ij}Z^{ij})g_E^{(0)\prime}.
\label{(79)}
\end{equation}
 From now on we will drop the 0 superscript and the primes on the
group elements with the understanding that the group elements are at
level zero and do not include the IGL(7) generators.
Equation \eqref{(79)} is easily evaluated using equations \eqref{(69a)}--\eqref{(69d)}, and we find that
\begin{equation}
E=(\det e)^{-{1\over 2}} \begin{pmatrix}
e_\mu{}^i& -{1\over \sqrt 2}e_\mu {}^j C_{j i_ii_2} \\
0&e^{-1}{}_{\mu_1\mu_2}{}^{i_1i_2} \end{pmatrix},
\label{(80)}
\end{equation}
where $e_\mu{}^i=(e^h)_\mu^i$,  and
$e^{-1}{}_{\mu_1\mu_2}{}^{i_1i_2}
=e^{-1}{}_{\mu_1}^{[i_1}e^{-1}{}_{\mu_2}{}^{i_2]}$. We are using
$\mu,\nu,\ldots $ as world indices in the four dimensional spacetime.
The prefactor follows from the terms with
${1\over 2}$ prefactors in equations \eqref{(69a)} and \eqref{(69b)},
which in turn were inherited from such terms in equations \eqref
{(38a)} and \eqref{(38b)}. As we
mentioned there, this precise prefactor arises from the fact that the
$l_1$ is a representation of $E_{11}$. We note that if one were to just
consider it as a ten dimensional representation  of SL(5) then the
factor
would be
$-{1\over 5}
$ rather than ${1\over 2}$, as we found in section \ref{2}.
\par
We will choose to construct the dynamics from the object defined in
equation
\eqref{(60)} which for simplicity we now denote by $M.$
Using equation \eqref{(80)} and $M=E E^{\#},$ where here $E^{\#} = E^T,$
\begin{equation}
M= (\det e )^{-1}
\begin{pmatrix} g_{\mu\nu}+{1\over 2} C_{\mu ij} C_{\nu}{}^{ij}& -
{1\over \sqrt 2} C_{\mu}{}^{\nu_1\nu_2} \\
-{1\over\sqrt
2}C_{\nu}{}^{\mu_1\mu_2}&g^{-1}{}^{[\mu_1|\nu_1|}g^{-1}{}^{\mu_2]\nu_2}
\end{pmatrix},
\label{(81)}
\end{equation}
where $C_{\mu ij}= e_\mu{}^k C_{k ij }$.
\par
The most general action which is quadratic in generalised spacetime
derivatives and invariant under the transformations of the non-linear
realisation was given in equation \eqref{(63)}. It involves five
constraints and
is unfamiliar in that it is defined over the extended space.
We now adopt the procedure explained at the end of section three.
Dropping the dependence of the fields on $x_{ij}$,  we now evaluate the
terms in the action of equation \eqref{(63)} to  find that
\begin{gather}
g^{-{1\over2}} M^{MN} (\partial_M M^{KL})( \partial_N M_{KL} ) =
3 g^{\mu \nu} (\partial_{\mu} g^{\sigma_1 \sigma_2 }) (\partial_{\nu}
g_{\sigma_1 \sigma_2}) - {{11}\over{2}}
g^{\mu \nu} (g^{\sigma_1 \sigma_2} \partial_{\mu} g_{\sigma_1
\sigma_2}) (g^{\tau_1 \tau_2}  \partial_{\nu} g_{\tau_1 \tau_2})
\notag \\
\hspace{65mm}- g^{\mu \nu} g^{\sigma_1 \dots \sigma_3, \tau_1 \dots \tau_3} (\partial_{\mu}
C_{\sigma_1 \dots \sigma_3})(\partial_{\nu} C_{\tau_1 \dots \tau_3}),
\label{(82)}
\end{gather}
\begin{align}
g^{-1/2} & M^{MN} (\partial_N M^{KL}) (\partial_L M_{MK}) \notag \\  &=
g^{\mu \sigma} (\partial_{\mu} g^{\nu \tau}) (\partial_{\nu} g_
{\sigma \tau}) -  (\partial_{\mu} g^{\mu \nu}) ( g^{\sigma \tau}
\partial_{\nu} g_{\sigma \tau})- {{1}\over{4}} g^{\mu \nu} (g^
{\sigma_1 \sigma_2} \partial_{\mu} g_{\sigma_1 \sigma_2}) (g^{\tau_1
\tau_2}  \partial_{\nu} g_{\tau_1 \tau_2})
\notag \\
& \hspace{65mm} -{{1}\over{2}} g^{\mu \tau_1} g^{\sigma_1 \sigma_2
\sigma_3, \nu \tau_2 \tau_3} (\partial_{\mu} C_{\sigma_1 \dots
\sigma_3})(\partial_{\nu} C_{\tau_1 \dots \tau_3}) ,
\label{(83)}
\end{align}
\begin{align}
g^{-1/2} & M_{RQ} \partial_{S} M^{SR} \partial_{P} M^{PQ} \notag \\
&=  g_{\sigma \tau}
(\partial_{\mu} g^{\mu \sigma}) (\partial_{\nu} g^{\nu \tau}) +
(\partial_{\mu} g^{\mu \nu})
( g^{\sigma \tau} \partial_{\nu} g_{\sigma \tau})
+ {1\over4} g^{\mu \nu} (g^{\sigma_1 \sigma_2} \partial_{\mu} g_
{\sigma_1 \sigma_2}) (g^{\tau_1 \tau_2}  \partial_{\nu} g_{\tau_1
\tau_2})  \notag \\
& \hspace{60mm} +{1\over2} g^{\mu \sigma_1} g^{\nu \tau_1} g^
{\sigma_2 \sigma_3, \tau_2 \tau_3}
(\partial_{\mu} C_{\sigma_1 \dots \sigma_3})(\partial_{\nu} C_
{ \tau_1 \dots \tau_3}),
\label{(84)}
\end{align}
\begin{equation}
g^{-1/2} M^{MN} (M^{KL} \partial_M M_{KL})(M^{RS} \partial_N M_{RS}) =
49 g^{\mu \nu} (g^{\sigma_1 \sigma_2} \partial_{\mu} g_{\sigma_1
\sigma_2}) (g^{\tau_1 \tau_2}  \partial_{\nu} g_{\tau_1 \tau_2}),
\label{(85)}
\end{equation}
and
\begin{equation}
g^{-1/2} \partial_{S} M^{ST} M^{PQ} \partial_{T} M^{PQ} =  -7
(\partial_{\mu} g^{\mu \nu})
( g^{\sigma \tau} \partial_{\nu} g_{\sigma \tau})
- {7\over2} g^{\mu \nu} (g^{\sigma_1 \sigma_2} \partial_{\mu} g_
{\sigma_1 \sigma_2}) (g^{\tau_1 \tau_2}  \partial_{\nu} g_{\tau_1
\tau_2}).
\label{(86)}
\end{equation}
Carrying out a gauge
transformation on the three form field we find the resulting action is
gauge invariant if $$ c_1= {1\over12}, \;\; c_2=-{1\over2}, \;\;
c_3=0, \;\; c_{4}={1\over84}, \;\;
c_5=0.$$ Up to integration by parts, the action is equal to
\begin{equation}
\int d^4 x \sqrt{g}(R-{{1}\over{48}} F^{(4)^{2}}),
\label{(87)}
\end{equation}
where $R$ is the Ricci scalar of the metric $g$ and $F^{(4)}$ is the
field strength of $C, F^{(4)}_{ijkl} = 4 \partial_{[i} C_{jkl]}.$ We
note that it is diffeomorphism invariant as well as U(1) gauge
invariant. After integration by parts the neglected boundary piece may be combined with the Gibbons-Hawking term to produce a boundary term for the generalised spacetime \cite{bermanmusaevperry}.

\section{Five dimensions: SO(5,5)}
\label{5}

\begin{figure}[http]
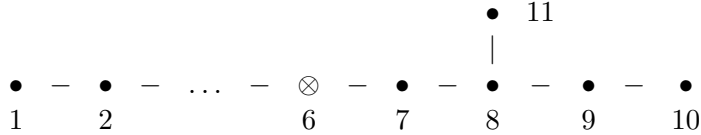

$$
\begin{array}{ccccccccccccccc}
     & &       & &         & &       & &       &  &\bullet&11&
& &        \\
     & &       & &       & &       & &       &  &   |   &  &
& &         \\
\bullet &-&\bullet&-&\ldots &-&\otimes&-&\bullet&- &\bullet&- &
\bullet&-& \bullet \\
    1    & &   2      & &         & &   6   & &   7   &  &   8   &
&   9   & &    10    \\
\end{array}
$$
\caption{The $E_{11}$ Dynkin diagram appropriate to the SO(5,5) duality}
\label{fig4}
\end{figure}

The non-linear realisation of the $E_{11} \ltimes l_{1}$ algebra will now be
used to find the generalised metric and construct an SO(5,5) duality
manifest dynamics, recovering the result of \cite{BGP} up to a
conformal factor. The conformal factor in \cite{BGP} was chosen to be a specific value. However, in this section, we see that the conformal factor is determined by the non-linear realisation of $E_{11} \ltimes l_{1}.$ The construction appropriate for the case of five
dimensions is found by deleting the sixth node of the $E_{11}$
Dynkin diagram, figure \ref{fig4}, to find the subalgebra GL(6) $
\otimes$ SO(5,5). As such we decompose $E_{11} \ltimes l_{1}$ into
representation of GL(6) $\otimes$ SO(5,5).

Consider the lowest level generators of $E_{11}$ given in equations
\eqref{(23)} and \eqref{(25)}. The generators that remain when we
truncate to five-dimensions are
$$  K^{i}_{\;j}, R^{ijk}, R_{ijk} \quad \textup{ and } \quad K^{a}{}_
{b},$$
where $i,j, \dots= 1, \dots, 5$ and $a,b, \dots = 1, \dots, 6.$
These generators are all at zero level, and the mixed index
generators are all at higher levels as in the case of SL(5). The
algebra that these generators satisfy is given by the truncation of
the $E_{11}$ algebra, equations \eqref{(24)}, \eqref{(26)}, \eqref{(28)} and \eqref{(32)}, to level zero
\begin{gather*}
[K^i{}_j, K^k{}_l]=\delta_j^k K^i{}_l-\delta_l^i K^k{}_j, \\
[K^i{}_j , R^{k_1k_2k_3}]= 3\delta^{[k_1}_j R^{|i|k_2k_3]} , \\
[K^i{}_j , R_{k_1k_2k_3}]=- 3\delta_{[k_1}^i R_{|j|k_2k_3]} , \\
[ R^{i_1i_2i_3}, R_{j_1j_2j_3} ]= 18 \delta^{[i_1i_2}_
{[j_1j_2} K^{i_3]}_{j_3 ]}-2\delta^{i_1 i_2 i_3}_
{j_1j_2j_3} (\sum_j K^j{}_j+\sum_a K^a{}_a)
; \\
[K^a{}_b, K^c{}_d ]=\delta^c_b K^a{}_d -
\delta^a_d K^c{}_b.
\end{gather*}
The $K^{a}_{\; b}$ clearly generate the GL(6) algebra, while $ {\tilde
K}^{i}_{\;j}, R^{ijk}, R_{ijk}$ generate the SO(5,5) algebra, where 
$$  {\tilde K}^{i}_{\;j} = K^{i}_{\;j} - {{1}\over{4}} \delta^{i}_{j}
\sum_{a} K^{a}{}_{a}.$$ 
We make the following identification
\[ M^{IJ}=
\begin{cases}
   {{1}\over{3!}} \epsilon^{IJklm} R_{klm} &\quad \textup{for } I,J=1,
\dots,5,\\
    \tilde{K}^{I}{}_{J-5} - {{1}\over{3}} \delta^{I}_{J-5} \sum_{k}
\tilde{K}^{k}{}_{k} &\quad \textup{for } I=1,\dots,5 \textup{ and }
J=6,\dots,10, \\
   {{1}\over{3!}} \epsilon_{(I-5)(J-5)klm} R^{klm} &\quad \textup
{for } I,J, \dots=6,\dots,10, \\
\end{cases}
\]
where $k,l,m= 1, \dots,5$ in the above. Now, one can see that the
generators $M^{IJ}$ satisfy the SO(5,5) algebra
$$ [M^{IJ}, M^{KL}] = \eta^{IK} M^{JL} - \eta^{IL} M^{JK} - \eta^{JK}
M^{IL} + \eta^{JL} M^{IK},$$
where
$$\eta= \begin{pmatrix}
          0 & 1_{5} \\
1_{5} &0
         \end{pmatrix}.
$$

Similarly taking the $l_1$ representation generators, given in
equation \eqref{(36)}, and restricting the indices to the case we are
considering, at the lowest level we find the generators
$$P_{i}, Z^{ij}, Z^{i_1 \dots i_5} \qquad \textup{and} \qquad P_{a},$$
where again $i,j, \dots= 1, \dots, 5$ and $a,b, \dots = 1, \dots, 6.
$  The first three generators generate the $\overline{16}$
representation of the SO(5,5) group, which we will call $\phi_
{\overline{16}},$ while $P_{a}$ generates translations in the 6-dimensional spacetime. The truncation of the $E_{11} \ltimes l_1$
algebra, equations \eqref{(37a)}--\eqref{(39e)}, gives the
commutation relations of these translation generators with the $E_{11}
$ generators
\begin{gather*}
[K^i{}_j, P_k]= -\delta _k^i P_j + {{1}\over{2}} \delta _j^i P_k, \\
[K^i{}_j, Z^{kl} ]= 2\delta_j^{[k} Z^{|i|l]}+{{1}\over{2}} \delta
_j^i Z^{kl}, \\
[K^i{}_j, Z^{k_1\ldots k_5} ]= 5\delta_j^{[k_1} Z^{|i|k_2\ldots k_5]}
+{{1}\over{2}}\delta _j^i Z^{k_1\ldots k_5} \\
[R^{i_1i_2i_3}, P_j]= 3 \delta^{[i_1}_j Z^{i_2i_3]}, \\
[R^{i_1i_2i_3}, Z^{jl} ]= Z^{i_1i_2i_3 jl},\\
[R^{i_1i_2i_3}, Z^{j_1\ldots j_5} ]=0,\\
[K^{a}{}_{b}, P_i]= {{1}\over{2}}\delta^{a}_{b} P_i, \\
[K^{a}{}_{b}, Z^{ij}]=  {{1}\over{2}} \delta^{a}_{b} Z^{kl}, \\
[K^{a}{}_{b}, Z^{i_1\ldots i_5} ]= {{1}\over{2}}\delta^{a}_{b} Z^{i_1
\ldots i_5}, \\
[K^i{}_j, P_{a}]= {{1}\over{2}} \delta _j^i P_{a}.
\end{gather*}
In what follows, we will use the Hodge dual of the $Z^{i_1 \dots i_5}
$ generator
$$W = {{1}\over{5!}} \epsilon_{i_1 \dots i_5} Z^{i_1 \dots i_5}$$ for
which the commutation relations can be easily found from the
commutations relations above.

As in the previous section, we will construct the non-linear
realisation using the SO(5,5) group element
$$g_{E} = \textup{e}^{h_{i}^{\;j} K^{i}_{\;j}} \textup{e}^{{{1}\over
{3!}} C_{ijk} R^{ijk}}, $$
which introduces the fields $h_{i}^{\;j} \textup{ and } C_{ijk}.$
Furthermore, the non-linear realisation also requires the group element
$$g_{l} = \textup{e}^{x^{i} P_{i}} \textup{e}^{{{1}\over{\sqrt{2}}} x_
{kl} Z^{kl}} \textup{e}^{w W },$$
which now has an extra generalised translation generator compared to
the SL(5) case. This introduces the coordinates $$ x^{i}, x_{kl}
\textup{ and } w,$$ which form the $16$ of SO(5,5). In the group
elements $g_{E}$ and $g_{l},$ we have, as before, left out the
generators $K^{a}{}_{b}$ and $P_{a}$, respectively. As in the
previous section this is a consistent truncation of generators. Using
the group elements $g_{E}$ and $g_{l}$ we construct the group element
of \eqref{(40)}
$$ g= \textup{e}^{x^{i} P_{i}+{{1}\over{\sqrt{2}}} x_{kl} Z^{kl}+ w W}
\textup{e}^{h_{i}^{\;j} K^{i}_{\;j}} \textup{e}^{{{1}\over{3!}} C_
{ijk} R^{ijk}} $$
from which the SO(5,5)$\ltimes \phi_{\overline{16}}$ non-linear
realisation can be constructed.

The non-linear realisation is carried out in a similar manner to that
outlined before, and ultimately one finds
\begin{align}
g^{-1}_{h} g^{-1}_{l} \textup{d}g_{l} g_{h} =& \textup{det}(\textup{e}
^h)^{-1/2}  (\textup{e}^h)_{j}^{\;\;i} \,\textup{d}x^{j} \left( P_{i}
- {{1}\over{2}} C_{ikl} Z^{kl} + {{1}\over{24}} C_{ik_1 k_2} C_{k_3
k_4 k_5} \epsilon^{k_1 \dots k_5} W \right) \notag \\
& \qquad + {{1}\over{\sqrt{2}}}  \textup{det}(\textup{e}^h)^{-1/2}
(\textup{e}^{-h})_{i}^{\;\;k} (\textup{e}^{-h})_{j}^{\;\;l} \, \textup
{d}x_{kl} \left( Z^{ij} - {{1}\over{6}} C_{k_1 k_2 k_3} \epsilon^
{ijk_1 k_2 k_3} W \right) \notag \\ & \qquad \qquad +  \textup{det}
(\textup{e}^h)^{-3/2}  \textup{d}w W.
\end{align}
The generalised vielbein can be read off from this expression,
\begin{equation}
E_{\Pi}{}^{A} = (\textup{det}e)^{-1/2}
\begin{pmatrix}
e_{\mu}{}^{i} & - {{1}\over{\sqrt{2}}} e_{\mu}{}^{j} C_{j i_1 i_2} &
{{1}\over{4}} e_{\mu}{}^{j} X_{j} \\
0 & e^{\mu_1}{}_{[i_1} e^{\mu_{2}}{}_{i_2]} & -{{1}\over{\sqrt{2}}} e^
{\mu_1}{}_{j_1} e^{\mu_{2}}{}_{j_2} V^{j_1 j_2} \\
0 &0& (\textup{det}e)^{-1}
\end{pmatrix},
\label{(89)}
\end{equation}
where $$V^{ij}={{1}\over{3!}} \epsilon^{ijklm}C_{klm} \qquad \textup
{and} \qquad X_{i}=C_{ijk} V^{jk}.$$
The tangent space indices are written with Latin letters and Greek
letters indicate space indices. We have also abbreviated the space
vielbein $\textup{e}^{h}$ to $e$ with the notation that $e_{\mu}{}^{i}
$ is the vielbein and $e^{\mu}{}_{i}$ is the inverse vielbein.

Hence the generalised metric, $M,$ for the SO(5,5) duality group is
\begin{equation}
M=g^{-1/2}
\begin{pmatrix} g_{\mu \nu}+{{1}\over{2}} C_{\mu}{}^{ij} C_{\nu ij} +
{{1}\over{16}}    X_{\mu}  X_{\nu}  &
{{1}\over{\sqrt{2}}} C_{\mu}{}^{ \nu_1 \nu_2} + {{1}\over{4 \sqrt
{2}}} X_{\mu} V^{\nu_1 \nu_2} & {{1}\over{4}} {g}^{-{1 / 2}} X_{\mu} \\
{{1}\over{\sqrt{2}}} C^{\mu_1 \mu_2}{}_{\nu} + {{1}\over{4 \sqrt
{2}}}  V^{\mu_1 \mu_2} X_{\nu}  & g^{\mu_1 \mu_2,\nu_1 \nu_2}+ {{1}
\over{2}} V^{\mu_1 \mu_2}V^{\nu_1 \nu_2} & {{1}\over{\sqrt{2}}}  g^{-
{1 / 2}} V^{\mu_1 \mu_2} \\
{{1}\over{4}} g^{-{1 / 2}} X_{\nu} & {{1}\over{\sqrt{2}}} g^{-{1 /
2}} V^{\nu_1 \nu_2} & g^{-1}
\end{pmatrix}
\label{(90)}
\end{equation}
where $g=(\textup{det}e)^{2}$ is the determinant of the metric $g_{\mu \nu}.$ This is the same generalised metric as in \cite{BGP}
except for the factor of $g.$ As we mentioned in section \ref{2}, and
shown in appendix \ref{B}, multiplying a metric by an overall factor
of $g$ does not change the fact that the generalised metric will
describe the dynamical theory. However, the factor of $g^{-1/2}$ in
the generalised metric, which one obtains by using the truncated $E_{11} \ltimes l_{1}$ algebra, will naturally lead to the incorporation
of the measure in the dynamics.

The generalised metric can now be used to describe the dynamics. The
following expression
\begin{gather*}
{{1}\over{16}} \, M^{MN} (\partial_M M^{KL})( \partial_N
M_{KL} ) - {{1}\over{2}} \, M^{MN} (\partial_N M^{KL}) (\partial_L M_
{MK}) \\
+ {{11}\over{1728}} \, M^{MN} (M^{KL} \partial_M M_{KL})(M^{RS}
\partial_N M_{RS}),
\end{gather*}
up to integration by parts, leads to the gauge-invariant and
diffeomorphism invariant combination
$$\sqrt{g} (R- {{1}\over{48}} F^{2}),$$
where $R$ is the Ricci scalar of the metric $g$ and $F= \textup{d} C$
is the field strength of the 3-form potential $C.$

\section{Six dimensions: $E_{6}$}
\label{6}

\begin{figure}[http]
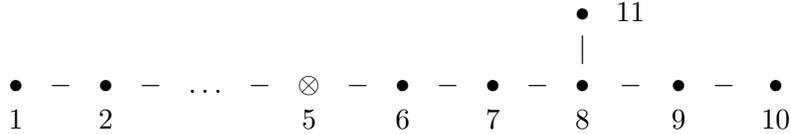

$$
\begin{array}{ccccccccccccccccc}
     & &       & &         & &       & &       &  &       &  &
\bullet&11&       & &        \\
     & &       & &       & &       & &       &  &       &  &   |   &
&          & &         \\
\bullet &-&\bullet&-&\ldots &-&\otimes&-&\bullet&- &\bullet&- &
\bullet&- &\bullet&-& \bullet \\
    1    & &   2      & &         & &   5   & &   6   &  &   7   &
&   8   &  &   9   & &    10    \\
\end{array}
$$
\caption{The $E_{11}$ Dynkin diagram appropriate to the $E_{6}$ duality}
\label{fig5}
\end{figure}

The non-linear realisation of $E_{11}\ltimes l_{1}$ for the case of
six dimensions to lowest level follows in the same way as before. We
begin by deleting the fifth node of $E_{11}$ Dynkin diagram, see
figure \ref{fig5}, to find the subalgebra appropriate to six
dimensions, GL(5) $\otimes E_{6}.$

Truncating the $E_{11}$ generators, equations \eqref{(23)} and \eqref
{(25)}, to the six dimensions at the lowest level, we find the group
generators
$$K^{i}_{\;j}, R^{ijk}, R_{ijk}, R^{i_{1} \dots i_{6}}, R_{i_{1}
\dots i_{6}} \qquad \textup{and} \qquad K^{a}{}_{b}, $$
where Latin letters from the middle of the alphabet $i,j, \dots= 1,
\dots, 6,$ and the start of the alphabet $a,b, \dots = 1, \dots, 5.$
These generators are those at level zero as before. The algebra
satisfied by these generators is found by truncating the $E_{11}$
algebra, equations \eqref{(24)}, \eqref{(26)}--\eqref{(28b)} and
\eqref{(30)}--\eqref{(33)}, appropriately, in which case we find the
algebra
\begin{gather*}
[K^i{}_j, K^k{}_l]=\delta_j^k K^i{}_l-\delta_l^i K^k{}_j, \\
[K^i{}_j , R^{k_1k_2k_3}]= 3\delta^{[k_1}_j R^{|i|k_2k_3]} , \\
[K^i{}_j , R_{k_1k_2k_3}]=- 3\delta_{[k_1}^i R_{|j|k_2k_3]} , \\
[K^i{}_j , R^{k_1 \dots k_6}]= 6\delta^{[k_1}_j R^{|i|k_2 \dots
k_6]} , \\
[K^i{}_j , R_{k_1 \dots k_6}]=- 6\delta_{[k_1}^i R_{|j|k_2\dots
k_6]} , \\
[ R^{i_1i_2i_3}, R^{j_1j_2j_3} ] = 2 R^{i_1i_2i_3j_1j_2j_3}, \\
[ R_{i_1i_2i_3}, R_{j_1j_2j_3} ] = 2 R_{i_1i_2i_3j_1j_2j_3}, \\
[ R^{i_1i_2i_3}, R_{j_1j_2j_3} ]= 18 \delta^{[i_1i_2}_
{[j_1j_2} K^{i_3]}_{j_3 ]}-2\delta^{i_1 i_2 i_3}_
{j_1j_2j_3} (\sum_j K^j{}_j+\sum_\mu K^\mu{}_\mu), \\
[ R^{i_1 \dots i_6}, R_{j_1 \dots j_6} ]= -5!.3.3 \, \delta^{[i_1
\dots i_5}_{[j_1 \dots j_5} K^{i_6]}_{j_6 ]} +5! \delta^{i_1 \dots
i_6}_{j_1 \dots j_6} (\sum_j K^j{}_j+\sum_a K^a{}_a), \\
[ R_{i_1i_2i_3}, R_{j_1 \dots j_6} ] = {{5!}\over{2}} \delta^{[j_1
j_2 j_3}_{i_1i_2i_3} R_{j_4j_5j_6]}; \\ [K^a{}_b , K^c{}_d ]=
\delta^c_b K^a{}_d - \delta^a_d K^c{}_b.
\end{gather*}
The $K^a{}_b$ generate the GL(5) algebra, while the generators
$$\tilde{K}^{i}{}_{j}= K^{i}{}_{j} -{{1}\over{3}} \delta^{i}_{j}
\sum_a K^a{}_a,$$
$R^{ijk}, R_{ijk}, R^{i_{1} \dots i_{6}}$ and $ R_{i_{1} \dots i_{6}}
$ generate the $E_{6}$ algebra.

The generalised translation generators can be found by considering
the generators of the $l_{1}$ representation of $E_{11},$ equation
\eqref{(36)}, at lowest level truncated to six dimensions. The
generators that we find in this case are
$$P_{i}, Z^{ij}, Z^{ijklm} \qquad \textup{and} \qquad P_{a}.$$ The
generators with indices labelled by Latin letters from the middle of
the alphabet generate the $\overline{27}$ representation of $E_{6},$
which we denote $\phi_{\overline{27}},$ while $P_{a}$ generates
translations along the extra 5 directions. From equations \eqref{(37a)}--\eqref{(39b)}, we can write down the commutation relations
for the translation generators, which are
\begin{gather}
[K^i{}_j, P_k]= -\delta _k^i P_j + {{1}\over{2}} \delta _j^i P_k,
\label{(91a)}\\
[K^i{}_j, Z^{kl} ]= 2\delta_j^{[k} Z^{|i|l]} +{{1}\over{2}} \delta
_j^i Z^{kl}, \label{(91b)} \\
[K^i{}_j, Z^{k_1\ldots k_5} ]= 5\delta_j^{[k_1} Z^{|i|k_2\ldots k_5]}
+{{1}\over{2}}\delta _j^i Z^{k_1\ldots k_5}, \label{(92)}\\
[R^{i_1i_2i_3}, P_j]= 3 \delta^{[i_1}_j Z^{i_2i_3]}, \label{(93a)}\\
[R^{i_1i_2i_3}, Z^{jl} ]= Z^{i_1i_2i_3 jl},\label{(93b)}\\
[R^{i_1i_2i_3}, Z^{j_1\ldots j_5} ]=0, \label{(93c)}\\
[R^{i_1 \dots i_6}, P_j]= - 3 \delta^{[i_1}_j Z^{i_2 \dots i_6]},
\label{(94a)}  \\
[R^{i_1 \dots i_6}, Z^{jl} ]= 0,\label{(94b)} \\
[R^{i_1 \dots i_6}, Z^{j_1\ldots j_5} ]=0, \label{(94c)} \\
[K^{a}{}_{b}, P_i]= {{1}\over{2}}\delta^{a}_{b} P_i, \\
[K^{a}{}_{b}, Z^{ij}]=  {{1}\over{2}} \delta^{a}_{b} Z^{kl}, \\
[K^{a}{}_{b}, Z^{i_1\ldots i_5} ]= {{1}\over{2}}\delta^{a}_{b} Z^{i_1
\ldots i_5}, \\
[K^i{}_j, P_{a}]= {{1}\over{2}} \delta _j^i P_{a}.
\end{gather}
For convenience, we will again use the Hodge dual of the $Z^{ijklm}$
generator $$W_{p}= {{1}\over{5!}}\epsilon_{pijklm} Z^{ijklm}.$$

Now, we are ready to construct the non-linear realisation of $ E_{6}
\ltimes \phi_{\overline{27}}.$ The group element of \eqref{(46)} is
$$g_{E} = \textup{e}^{h_{i}^{\;j} K^{i}_{\;j}} \textup{e}^{{{1}\over
{3!}} C_{ijk} R^{ijk}} \textup{e}^{{{1}\over{6!}} C_{i_1 \dots i_6} R^
{i_1 \dots i_6}},$$
which introduces the fields
$$ h_{i}^{\;j},C_{ijk} \textup{  and  } C_{i_1 \dots i_6}.$$ Note
that in six dimensions a new field $C_{i_1 \dots i_6},$ which is a 6-form potential, is introduced. This was not present in previous
examples because in those cases the dimensions we were considering
were less than six.
Further to the group element, $g_{E},$ there is the group element
$$g_{l} = \textup{e}^{x^{i} P_{i}} \textup{e}^{{{1}\over{\sqrt{2}}} x_
{kl} Z^{kl}} \textup{e}^{w^{i} W_{i} },$$
which introduces the coordinates
$$x^{i}, x_{kl} \textup{ and } w^{i}.$$ These form the 27 of $E_{6}.$
It is again consistent to leave out the generators $K^{a}{}_{b}$ and
$P_{a}$ from the non-linear realisation.

We now calculate the Maurer-Cartan form for the non-linear
realisation and hence the generalised vielbein, equation \eqref
{(53)}. By Hodge dualising equation \eqref{(92)}, we can find that $$
[K^{i}_{\;\;j}, W_{k}]= - \delta^{i}_{k} W_{j} +{{3}\over{2}} \delta^
{i}_{j} W_{k}.$$ Now, using the above commutation relation and
equations \eqref{(91a)} and \eqref{(91b)}, we conjugate the Maurer-Cartan form of $g_{l}$ by $\textup{e}^{h_{i}^{\;j} K^{i}_{\;j}}$ to
obtain
\begin{align}
\textup{e}^{-h_{i}^{\;j} K^{i}_{\;j}} g_{l}^{-1} \textup{d}g_{l}
\textup{e}^{h_{k}^{\;l} K^{k}_{\;l}} &= \textup{det}(\textup{e}^h)^{-
{{1}\over{2}}} \left( (\textup{e}^h)_{\mu}{}^{i} \,\textup{d}x^{\mu}
P_{i} + \textstyle{{{1}\over{\sqrt{2}}}} (\textup{e}^{-h})_{i}{}^
{\mu} (\textup{e}^{-h})_{j}{}^{\nu} \, \textup{d}x_{\mu \nu} Z^{ij}
\right. \notag \\
& \hspace{65mm}   \left.
+ \textup{det}(\textup{e}^h)^{-1} (\textup{e}^{h})_{\mu}{}^{i} \textup
{d}w^{\mu} W_{i} \right),
\label{(96)}
\end{align}
where Greek and Latin letters denote spacetime and tangent space
indices, respectively.
This gives the dependence of the generalised vielbein on the
spacetime metric, and conjugating the above expression by $\textup{e}^
{{{1}\over{3!}} C_{ijk} R^{ijk}}$ we obtain the dependence on the 3-form potential:
\begin{align}
&\textup{e}^{-{{1}\over{3!}} C_{ijk} R^{ijk}} \textup{e}^{-h_{i}^
{\;j} K^{i}_{\;j}} g_{l}^{-1} \textup{d}g_{l} \textup{e}^{h_{k}^{\;l}
K^{k}_{\;l}} \textup{e}^{{{1}\over{3!}} C_{ijk} R^{ijk}} \notag \\
=& \, \textup{det}(\textup{e}^h)^{-1/2} (\textup{e}^h)_{\mu}{}^{i} \,
\textup{d}x^{\mu} P_{i}
+ {{1}\over{\sqrt{2}}} \textup{det}(\textup{e}^h)^{-1/2} (\textup{e}^
{-h})_{i}{}^{\mu} (\textup{e}^{-h})_{j}{}^{\nu} \left( \textup{d} x_
{\mu \nu} - {{1}\over{\sqrt{2}}} C_{\mu \nu \rho} \textup{d}x^{\rho}
\right) Z^{ij} \notag \\  & \qquad \qquad  + \textup{det}(\textup{e}
^h)^{-3/2} (\textup{e}^{h})_{\mu}{}^{i} \left( \textup{d}w^{\mu} -
{{1}\over{\sqrt{2}}} V^{\mu \nu \rho} \textup{d}x_{\nu \rho} + {{1}
\over{4}} C_{\nu kl} V^{\mu kl} \textup{d}x^{\nu} \right) W_{i},
\label{(97)}
\end{align}
where we have defined $V^{ijk}= {{1}\over{3!}} \epsilon^{ijklmn} C_
{lmn}.$ In the deriving the above expression we have made use of
equation \eqref{(93a)} and a rewriting of equation \eqref{(93b)}, $$
[R^{ijk}, Z^{mn}]= \epsilon^{pijkmn} W_{p}.$$ Note that in the
truncation to six-dimensions the commutator of $R^{ijk}$ with $Z^{i_1
\dots i_5}$ is zero because $Z^{i_{1} \dots l_{7} , a}$ vanishes.

Finally, we conjugate by the group element given by exponentiation of
the $R^{i_{1} \dots i_{6}}$ generator. Note that the only non-vanishing commutation relation of $R^{i_{1} \dots i_{6}}$ with a
generalised translation generator is the commutation relation with $P_
{j},$ equation \eqref{(94a)}, or equivalently $$[R^{i_{1} \dots i_
{6}}, P_{d}] = 3 \, \delta^{[i_1}_{d} \epsilon^{i_1 \dots i_6 ] p} W_
{p}.$$ This gives us the dependence of the generalised vielbein on
the 6-form potential. All in all, we obtain
\begin{align}
g_{E}^{-1} g_{l}^{-1} \textup{d}g_{l} g_{E} &= \, \textup{det}(\textup
{e}^h)^{-1/2} (\textup{e}^h)_{\mu}{}^{i} \,\textup{d}x^{\mu} P_{i}
\notag \\
& \quad  + {{1}\over{\sqrt{2}}} \textup{det}(\textup{e}^h)^{-1/2}
(\textup{e}^{-h})_{i}{}^{\mu} (\textup{e}^{-h})_{j}{}^{\nu} \left
( \textup{d} x_{\mu \nu} - {{1}\over{\sqrt{2}}} C_{\mu \nu \rho}
\textup{d}x^{\rho} \right) Z^{ij} \notag \\
&  \qquad + \textup{det}(\textup{e}^h)^{-3/2} (\textup{e}^{h})_{\mu}{}
^{i} \left( \textup{d}w^{\mu} - {{1}\over{\sqrt{2}}} \textup{det}
(\textup{e}^h) V^{\mu \nu \rho} \textup{d}x_{\nu \rho} \right. \notag \\
& \hspace{45mm} \left. + {{1}\over{4}} \textup{det}(\textup{e}^h) C_
{kl\nu} V^{\mu kl} \textup{d}x^{\nu} + {{1}\over{2}} \textup{det}
(\textup{e}^h) U \textup{d}x^{\mu} \right) W_{i},
\label{(98)}
\end{align}
where $U$ is the Hodge dual of the 6-form potential, $$U= {{1}\over
{6!}} \epsilon^{i_1 \dots i_6} C_{i_1 \dots i_6}.$$

Now, we can read off the generalised vielbein from equation \eqref{(98)}. Using the same notation as before for the ordinary space
vielbein, the generalised vielbein is
\begin{equation}
E_{\Pi}{}^{A} =  (\textup{det}e)^{-1/2}
\begin{pmatrix}
e_{\mu}{}^{i} & - {{1}\over{\sqrt{2}}} e_{\mu}{}^{j} C_{j i_1 i_2} &
{{1}\over{2}} e_{\mu}{}^{i_3} U + {{1}\over{4}} e_{\nu}{}^{i_3} C_
{\mu j k} V^{\nu j k} \\
0 & e^{\mu_1}{}_{[i_1} e^{\mu_{2}}{}_{i_2]} & -{{1}\over{\sqrt{2}}} e^
{\mu_1}{}_{j_1} e^{\mu_{2}}{}_{j_2} V^{j_1 j_2 i_3} \\
0 & 0 & (\textup{det}e)^{-1} e_{\mu_3}^{i_3}
\end{pmatrix}.
\label{(99)}
\end{equation}
This generalised vielbein is very similar to the generalised vielbein
in the case of the SO(5,5) duality group. In fact the metric and 3-form potential dependence of the two generalised vielbein are
identical, except for the obvious difference that third generalised
coordinate direction in this case has an index but this is only
because we are using the Hodge dual of $Z^{i_1 \dots i_5}.$ The
dependence of the generalised vielbein on the 3-form potential only
changes when there is a new generalised coordinate direction in which
case higher order terms in the 3-form potential enter the generalised
vielbein. In contrast, though, the generalised vielbein for the $E_{6}
$ duality group gives the dependence of the generalised vielbein on
the 6-form potential.

The generalised metric corresponding to the generalised vielbein,
expression \eqref{(99)}, is constructed using equation \eqref{(60)}.
The dynamics can then be written in terms of this generalised metric.
The combination
\begin{multline}
{{1}\over{24}} \, M^{MN} (\partial_M M^{KL})( \partial_N
M_{KL} ) -{{1}\over{2}} \, M^{MN} (\partial_N M^{KL}) (\partial_L M_
{MK}) \\
+ {{19}\over{9720}} \, M^{MN} (M^{KL} \partial_M M_{KL})(M^{RS}
\partial_N M_{RS}),
\label{(100)}
\end{multline}
again, up to integration by parts, reproduces $$\sqrt{g} \left(R -
{{1}\over{48}} {F^{(4)}}^{2}\right)$$ when derivatives with respect
to the extra generalised coordinates are taken to vanish.

The 6-form potential is not dynamical in 6-dimensions as its gauge-invariant field strength vanishes. When one evaluates the expression in \eqref{(100)} one discovers that $C^{(6)}$ cancels completely, verifying that the 6-form potential does not contribute to the action.

\section{Seven dimensions: $E_{7}$}
\label{7}

\begin{figure}[http]
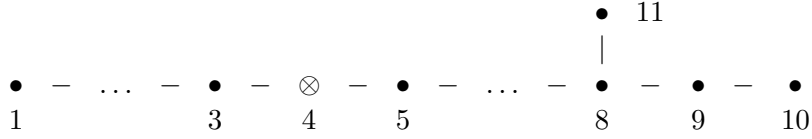

$$
\begin{array}{ccccccccccccccccc}
     & &       & &         & &       & &       &  &       &  &
\bullet&11&       & &        \\
     & &       & &       & &       & &       &  &       &  &   |   &
&          & &         \\
\bullet &-&\ldots &-&\bullet&-&\otimes&-&\bullet&- &\ldots &- &
\bullet&- &\bullet&-& \bullet \\
    1    & &       & &     3   & &   4   & &   5   &  &       &  &
8   &  &   9   & &    10    \\
\end{array}
$$
\caption{The $E_{11}$ Dynkin diagram appropriate to the $E_{7}$ duality}
\label{fig6}
\end{figure}

In this section, we apply the non-linear realisation of $E_{11}
\ltimes l_{1} $ to seven dimensions. This is found by deleting the
fourth node of the $E_{11}$ Dynkin diagram, see figure \ref{fig6}, in
which case we find the subalgebra GL(4)$\otimes E_{7}.$

The $E_{11}$ algebra of generators, equations \eqref{(23)} and \eqref
{(25)}, at level zero with respect to the deletion of node four are
$$K^{i}_{\;j}, R^{ijk}, R_{ijk}, R^{i_{1} \dots i_{6}}, R_{i_{1}
\dots i_{6}} \qquad \textup{and} \qquad K^{a}{}_{b}, $$
where the indices labelled $i, j, \dots $ run from 1 to 7, while
those labelled by $a, b, \dots $ run from 1 to 4. The commutation
relations between these generators can be read off from the $E_{11}$
algebra, equations \eqref{(24)}, \eqref{(26)}--\eqref{(28b)} and
\eqref{(30)}--\eqref{(33)},
\begin{gather*}
[K^i{}_j, K^k{}_l]=\delta_j^k K^i{}_l-\delta_l^i K^k{}_j, \\
[K^i{}_j , R^{k_1k_2k_3}]= 3\delta^{[k_1}_j R^{|i|k_2k_3]} , \quad
[K^i{}_j , R_{k_1k_2k_3}]=- 3\delta_{[k_1}^i R_{|j|k_2k_3]} , \\
[K^i{}_j , R^{k_1 \dots k_6}]= 6\delta^{[k_1}_j R^{|i|k_2 \dots
k_6]} , \quad
[K^i{}_j , R_{k_1 \dots k_6}]=- 6\delta_{[k_1}^i R_{|j|k_2\dots
k_6]} , \\
[ R^{i_1i_2i_3}, R^{j_1j_2j_3} ] = 2 R^{i_1i_2i_3j_1j_2j_3}, \quad
[ R_{i_1i_2i_3}, R_{j_1j_2j_3} ] = 2 R_{i_1i_2i_3j_1j_2j_3}, \\
[ R^{i_1i_2i_3}, R_{j_1j_2j_3} ]= 18 \delta^{[i_1i_2}_
{[j_1j_2} K^{i_3]}{}_{j_3 ]}-2\delta^{i_1 i_2 i_3}_
{j_1j_2j_3} (\sum_j K^j{}_j+\sum_a K^a{}_a), \\
[ R^{i_1 \dots i_6}, R_{j_1 \dots j_6} ]= -5!.3.3 \, \delta^{[i_1
\dots i_5}_{[j_1 \dots j_5} K^{i_6]}{}_{j_6 ]} +5! \delta^{i_1 \dots
i_6}_{j_1 \dots j_6} (\sum_j K^j{}_j+\sum_a K^a{}_a); \\
[ R_{i_1i_2i_3}, R^{j_1 \dots j_6} ] = {{5!}\over{2}} \delta^{[j_1
j_2 j_3}_{i_1i_2i_3} R^{j_4j_5j_6]}, \\ [K^a{}_b , K^c{}_d ]=
\delta^c_b K^a{}_d -
\delta^a_d K^c{}_b.
\end{gather*}
The $E_{7}$ algebra derived from Cartan's 56-dimensional
representation of $E_{7}$ \cite{cartanthese, cartan, freudenthal},
see appendix \ref{C}, can be recovered from these relations by
shifting the GL(7) generator, $K^{i}{}_{j},$ by the trace of the GL(4) generators $K^{a}{}_{b}$
$$ \tilde{K}^{i}{}_{j} = K^{i}{}_{j} - {{1}\over{2}} \delta^{i}{}_{j}
\sum_{a} K^{a}{}_{a}.$$

The list of $l_{1}$ generators, equation \eqref{(36)}, can similarly
be truncated to seven dimensions where we find the generators
$$P_{i}, Z^{ij}, Z^{i_1 \dots i_5}, Z^{i_1 \dots i_7,j} \quad \textup
{ and } \quad P_{a}.$$
The first four generate the $56$ representation of $E_{7},$ denoted $\phi_{56},$ and $P_{a}$ generate translations along the four extra
directions. The $E_{11} \ltimes l_{1}$ algebra, equations \eqref{(37a)}--\eqref{(39b)} gives the commutation relations of the
generalised translation generators with the GL(4)$\otimes E_{7}$
generators. For convenience, we will use the generators
\begin{equation}
W_{ij} = {{1}\over{5!}}\epsilon_{ij  k_1 \dots k_5}  Z^{k_1 \dots
k_5}, \qquad W^{i} = {{1}\over{7!}} \epsilon_{j_1 \dots j_7} Z^{j_1
\dots j_7,i},
\end{equation}
and write the commutation relations in terms of these generators.
\begin{gather*}
[ K^{i}_{\;j} , P_{k} ] = - \delta^{i}_{k} P_{j} + {{1}\over{2}}
\delta^{i}_{j} P_{k}, \qquad
[ K^{i}_{\;j} , Z^{kl} ] = 2 \, \delta^{[k}_{j} Z^{|i|l]} + {{1}\over
{2}} \delta^{i}_{j} Z^{kl},  \\
[ K^{i}_{\;j} , W_{kl} ] = -2 \, \delta^{i}_{[k} W_{|j| l ]} + {{3}
\over{2}} \delta^{i}_{j} W_{kl}, \qquad
[ K^{i}_{\;j} , W^{k} ] =  \delta^{k}_{j} W^{i} + {{3}\over{2}}
\delta^{i}_{j} W^{k},\\
[ R_{ijk} , P_{l} ] = 0, \qquad
[ R_{ijk} , Z^{mn} ] = 3! \, \delta^{mn}_{[ij} P_{k]},  \\
[ R_{ijk} , W_{mn}  ] = {{1}\over{2}} \epsilon_{ijkmnpq} Z^{pq}, \qquad
[ R_{ijk} , W^{l} ] = {{1}\over{560}} \delta^{l}_{[i } W_{jk]}, \\
[ R^{ijk} , P_{l} ] = 3 \, \delta^{[i}_{l} Z^{jk]}, \qquad
[ R^{ijk} , Z^{mn} ] = {{1}\over{2}} \epsilon^{ijkmnpq} W_{pq}, \\
[ R^{ijk} , W_{mn} ] = 2 \, \delta^{[ij}_{mn} W^{k]}, \qquad
[ R^{ijk} , W^{l} ] = 0, \\
[ R_{i_1 \dots i_6}, P_{j}] = 0, \qquad
[ R_{i_1 \dots i_6}, Z^{kl}] = 0, \\
[ R_{i_1 \dots i_6}, W_{kl} ] = -3 \epsilon_{kl [i_1 \dots i_5 } P_
{i_6]}, \qquad
[ R_{i_1 \dots i_6}, Z^{j}] = -{{3}\over{2}} \epsilon_{k i_1 \dots
i_6} Z^{kj}, \\
[ R^{i_1 \dots i_6}, P_{j}] = {{1}\over{2}} \epsilon^{i_1 \dots i_6
k} W_{jk}, \qquad
[ R^{i_1 \dots i_6}, Z^{kl}] = {{1}\over{3}} \epsilon^{i_1 \dots i_6
[k} W^{l]},\\
[ R^{i_1 \dots i_6}, W_{kl}] = 0, \qquad
[ R^{i_1 \dots i_6}, W^{j} ] = 0, \\
[K^{a}{}_{b}, P_i]= {{1}\over{2}}\delta^{a}_{b} P_i, \qquad  [K^{a}{}_
{b}, Z^{ij}]=  {{1}\over{2}} \delta^{a}_{b} Z^{kl},
\\ [K^{a}{}_{b}, W_{ij} ]= {{1}\over{2}} \delta^{a}_{b} W_{ij},
\qquad [K^{a}{}_{b}, W^{i}]= {{1}\over{2}}\delta^{a}_{b} W^{i} \qquad
[K^i{}_j, P_{a}]= {{1}\over{2}} \delta _j^i P_{a}.
\end{gather*}
In appendix \ref{C}, we show that the generators $$\tilde{K}^{i}_
{\;j}, R^{ijk}, R_{ijk},R^{i_1 \dots i_6},R_{i_1 \dots i_6} \quad
{\rm and } \quad P_{i}, Z^{ij}, Z^{i_1 \dots i_5}, Z^{i_1 \dots i_7,j}
$$ do indeed generate the $E_{7} \ltimes \phi_{56}$ algebra.

We can now construct the non-linear realisation, equation \eqref{(40)}, for $E_{7} \ltimes \phi_{56}$ and find the generalised
metric. The objects from which the non-linear realisation is
constructed are the group element, equation \eqref{(41)},
$$g_{E} = \textup{e}^{h_{i}^{\;j} K^{i}_{\;j}} \textup{e}^{{{1}\over
{3!}} C_{ijk} R^{ijk}} \textup{e}^{{{1}\over{6!}} C_{i_1 \dots i_6} R^
{i_1 \dots i_6}},$$
which introduces the fields
$$ h_{i}^{\;j},C_{ijk} \textup{ and } C_{i_1 \dots i_6},$$
and the group element, equation \eqref{(42)},
$$g_{l} = \textup{e}^{x^{i} P_{i}} \textup{e}^{{{1}\over{\sqrt{2}}} x_
{ij} Z^{ij}} \textup{e}^{{{1}\over{\sqrt{2}}} w^{ij} W_{ij}} \textup
{e}^{{{1}\over{3}} w_{i} W^{i} },$$
which introduces the generalised coordinates
$$ x^{i}, x_{kl}, w^{kl} \textup{ and } w_{i}.$$ The generalised
coordinates are in the $\overline{56}$ of $E_{7}.$

Now, the generalised vielbein is constructed from $$g_{E}^{-1} g_{l}^{-1} \textup{d}g_{l} g_{E}.$$ Similar calculation to the calculations
in the previous sections show that the generalised vielbein, $E_{\Pi}{}^{A},$ is
{\small
\begin{equation}
\textup{e}^{-{{1}\over{2}}}
\begin{pmatrix}
e_{\mu}{}^{i} & - {{1}\over{\sqrt{2}}} e_{\mu}{}^{j} C_{j i_1 i_2} &
{{1}\over{\sqrt{2}}} e_{\mu}{}^{[i_3} U^{i_4]} + {{1}\over{4\sqrt
{2}}}  e_{\mu}{}^{j} X_{j;}{}^{i_3 i_4} & {{1}\over{2}} e_{\mu}{}^{j}
C_{j i_5 k }  U^{k} - {{1}\over{24}} e_{\mu}{}^{j} X_{j;}{}^{kl} C_
{kl i_5}\\
0 & e^{\mu_1}{}_{[i_1} e^{\mu_{2}}{}_{i_2]} & -{{1}\over{\sqrt{2}}} e^
{\mu_1}{}_{j_1} e^{\mu_{2}}{}_{j_2} V^{j_1 j_2 i_3 i_4} & {{1}\over
{\sqrt{2}}} e^{\mu_1}{}_{[j} e^{\mu_{2}}{}_{i_5]} U^{j} + {{1}\over{4
\sqrt{2}}} e^{\mu_1}{}_{j_1} e^{\mu_{2}}{}_{j_2} X_{i_5 ;}{}^{j_1
j_2} \\
0 & 0 & \textup{e}^{-1} e_{\mu_3}{}^{[i_3} e_{\mu_{4}}{}^{i_4]} & -
{{1}\over{\sqrt{2}}} e_{\mu_3}{}^{j_1} e_{\mu_{4}}{}^{j_2} C_{j_1 j_2
i_5} \\
0 & 0 & 0 & \textup{e}^{-1} e^{\mu_5}{}_{i_5} \\
\end{pmatrix},
\label{(102)}
\end{equation}}
where $\textup{e}$ is the determinant of the vielbein $e$ and $$ g_{\mu \nu} = e^{i}_{\mu} e^{j}_{\nu} \eta_{ij};$$ $V^{i_1 \dots i_4},
U^{i}$ are Hodge duals of the 3-form and 6-form potentials,
respectively, $$V^{i_1 \dots i_4}= {{1}\over{3!}} \epsilon^{i_1 \dots
i_4 j_1 \dots j_3} C_{j_1 \dots j_3}, \qquad U^{i}= {{1}\over{6!}}
\epsilon^{i j_1 \dots j_6} C_{j_1 \dots j_6};$$ and $$X_{i;}{}^{jk}=
C_{ilm} V^{jklm}.$$ The indices labelled by Greek indices in the
expression for the generalised vielbein are tangent space indices and
Latin letters label space indices.

We can find that when we restrict the fields to only depend on
ordinary space coordinates then
\begin{align}
   g^{-1/2} M^{MN} (\partial_M M^{KL})( \partial_N M_{KL} )  =& 12 g^
{\mu \nu} (\partial_{\mu} g^{\sigma \tau}) (\partial_{\nu} g_{\sigma
\tau}) -62 g^{\mu \nu} (g^{\sigma_1 \sigma_2} \partial_{\mu} g_
{\sigma_1 \sigma_2}) (g^{\tau_1 \tau_2}  \partial_{\nu} g_{\tau_1
\tau_2}) \notag \\
&  \quad  -4 g^{\mu \nu} g^{\sigma_1 \dots \sigma_3, \tau_1 \dots
\tau_3} (\partial_{\mu} C_{\sigma_1 \dots \sigma_3})(\partial_{\nu} C_
{\tau_1 \dots \tau_3}) \notag \\
&\qquad   - {{1}\over{5!}} g^{\mu \nu} g^{\sigma_1 \dots \sigma_6,
\tau_1 \dots \tau_6} (\partial_{\mu} C_{\sigma_1 \dots \sigma_6} - 20
C_{[\sigma_1 \dots \sigma_3|} \partial_{\mu} C_{|\sigma_4 \dots
\sigma_6]} ) \notag \\
&\qquad \qquad \qquad \;  \times (\partial_{\nu} C_{\tau_1 \dots
\tau_6} - 20 C_{[\tau_1 \dots \tau_3|} \partial_{\nu} C_{|\tau_4
\dots \tau_6]}),
\label{(103)}
\end{align}
\begin{align}
   g^{-1/2} M^{MN} (\partial_N M^{KL}) (\partial_L M_{MK})  &=  g^
{\mu \sigma} (\partial_{\mu} g^{\nu \tau}) (\partial_{\nu} g_{\sigma
\tau}) -  (\partial_{\mu} g^{\mu \nu}) ( g^{\sigma \tau} \partial_
{\nu} g_{\sigma \tau}) \notag\\
& \; - {{1}\over{4}} g^{\mu \nu} (g^{\sigma_1 \sigma_2} \partial_
{\mu} g_{\sigma_1 \sigma_2}) (g^{\tau_1 \tau_2}  \partial_{\nu} g_
{\tau_1 \tau_2}) \notag \\
& \;\;-{{1}\over{2}} g^{\mu \tau_1} g^{\sigma_1 \sigma_2 \sigma_3,
\nu \tau_2 \tau_3} (\partial_{\mu} C_{\sigma_1 \dots \sigma_3})
(\partial_{\nu} C_{\tau_1 \dots \tau_3}) \notag\\
&\;\;\;  - {{1}\over{4(5!)}} g^{\mu \tau_1} g^{\sigma_1 \dots
\sigma_6,  \nu \sigma_2 \dots \sigma_6} (\partial_{\mu} C_{\sigma_1
\dots \sigma_6} - 20 C_{[\sigma_1 \dots \sigma_3|} \partial_{\mu} C_{|
\sigma_4 \dots \sigma_6]} ) \notag \\
& \qquad \qquad \qquad \quad \times (\partial_{\nu} C_{\tau_1 \dots
\tau_6} - 20 C_{[\tau_1 \dots \tau_3|} \partial_{\nu} C_{|\tau_4
\dots \tau_6]}),
\label{(104)}
\end{align}
\begin{align}
    g^{-1/2} M^{MN} (M^{KL}  \partial_M M_{KL})(M^{RS} \partial_N M_
{RS}) = 56^2 g^{\mu \nu} (g^{\sigma_1 \sigma_2} \partial_{\mu} g_
{\sigma_1 \sigma_2}) (g^{\tau_1 \tau_2}  \partial_{\nu} g_{\tau_1
\tau_2}).
\label{(105)}
\end{align}
In the above calculations we made use of the following identities
\begin{align*}
C_{\nu \sigma_1 \sigma_2} \partial_{\mu} V^{\sigma_1 \sigma_2 \tau_1
\tau_2} =  V^{\sigma_1 \sigma_2 \tau_1 \tau_2} \partial_{\mu} C_{\nu
\sigma_1 \sigma_2} + {{2}\over{3}} \delta^{[\tau_1} _{\nu} V^{\tau_2]
\sigma_1 \dots \sigma_3} \partial_{\mu} C_{\sigma_1 \dots \sigma_3}
- {{1}\over{2}} X_{\nu;}{}^{\tau_1 \tau_2} g^{\sigma_1 \sigma_2}
\partial_{\mu} g_{\sigma_1 \sigma_2},
\end{align*}
$$
C_{\sigma \mu_1 \mu_2} V^{\sigma \nu_1 \dots \nu_3} =  {{3}\over{2}}
\delta^{[\nu_1}_{[\mu_1} X_{\mu_2];}{}^{\nu_2 \nu_3]}
$$
which can be proved by Hodge dualising $C$ and $V$ and then
contracting the epsilon tensors. It is also useful to note that
\begin{align*}
C_{\nu_1 \dots \nu_3} V^{\mu \nu_1 \dots \nu_3} =  {{1}\over{3!}}
\epsilon^{\mu \nu_1 \dots \nu_3 \sigma_1 \dots \sigma_3}  C_{\nu_1
\dots \nu_3}  C_{\sigma_1 \dots \sigma_3}
\end{align*}
vanishes because the epsilon tensor makes exchanging the set of
indices $\nu_1 \dots \nu_3 $ and $\sigma_1 \dots \sigma_3 $ an
antisymmetric operation.

Now, in equations \eqref{(103)}--\eqref{(105)}, comparing the terms
that lead to the Ricci scalar, which is
\begin{align}
R &= {{1}\over{4}} g^{\mu \nu} (\partial_{\mu} g^{\sigma \tau})
(\partial_{\nu} g_{\sigma \tau}) - {{1}\over{2}} g^{\mu \sigma}
(\partial_{\mu} g^{\nu \tau}) (\partial_{\nu} g_{\sigma \tau}) \notag \\
&\hspace{40mm}  + {{1}\over{2}}  (\partial_{\mu} g^{\mu \nu}) ( g^
{\sigma \tau} \partial_{\nu} g_{\sigma \tau}) + {{1}\over{4}} g^{\mu
\nu} (g^{\sigma_1 \sigma_2} \partial_{\mu} g_{\sigma_1 \sigma_2}) (g^
{\tau_1 \tau_2}  \partial_{\nu} g_{\tau_1 \tau_2})
\label{(106)}
\end{align}
up to terms that are total derivatives, we conclude that the combination
\begin{multline}
{{1}\over{48}} \, M^{MN} (\partial_M M^{KL})( \partial_N
M_{KL} ) - {{1}\over{2}} \, M^{MN} (\partial_N M^{KL}) (\partial_L M_
{MK}) \\
+ {{17}\over{37632}} \, M^{MN} (M^{KL} \partial_M M_{KL})(M^{RS}
\partial_N M_{RS})
\label{(107)}
\end{multline}
leads to the Ricci scalar. In fact, when the fields are allowed to
only depend on ordinary space directions, this  reduces, up to
integration by parts, to
$$\sqrt{g} \left( R - {{1}\over{48}} {F^{(4)}}^{2} - {{1}\over{8!}}
{F^{(7)}}^{2} \right),$$
where $\sqrt{g} $ is the measure, $F^{(4)}$ is the field strength of
the 3-form potential,
$$ F^{(4)}_{\mu_1 \dots \mu_4} = 4 \partial_{[\mu_1} C_{\mu_2 \dots
\mu_4]}, $$ and $F^{(7)}$ is the field strength of 6-form potential,
$$ F^{(7)}_{\mu_1 \dots \mu_7} = 7 \partial_{[\mu_1} C_{\mu_2 \dots
\mu_7]} + 140 C_{[\mu_1 \dots \mu_3} \partial_{\mu_4} C_{\mu_5 \dots
\mu_7]}.$$
In the full theory in eleven dimensions one knows that the four and
seven form field strengths are dual. However, here we are considering
the theory in seven dimensions, so we cannot find an eleven-dimensional duality relation. The duality relation between these
fields should be recovered if one carries out the  non-linear
realisation of
$E_{11}\ltimes l_1$ in eleven dimensions. If one included \textit{all} the components of $h, C^{(3)}, C^{(6)}$ rather than just those
where one has $E_{7}$ indices, then one expects to be able to
reproduce the duality relation between $F^{(4)}$ and $F^{(7)}$.
Indeed, $E_{11}\ltimes l_1$ contains all the fields required to have
equations of motion that are only first order in spacetime derivatives.

The generalised vielbein, expression \eqref{(102)}, is the same as
that found in \cite{hillmann}, up to factors of $\textup{det } e$. In \cite{hillmann}, the dynamics is constructed in a different way and the other GL(4) directions are needed in order to construct the action. However, here  we formulate the dynamics using the generalised metric and find that imposing gauge
invariance automatically results in the action that is invariant
under diffeomorphisms, and vice-versa.

\paragraph*{Acknowledgements}
We would like to thank Gary Gibbons, Mahdi Godazgar, Hugh Osborn and Antony Wassermann for discussions. DSB is supported in part by the Queen Mary STFC rolling grant ST/G000565/1. HG is supported by an STFC grant and thanks St.\ John's College Cambridge for their support. MJP is in part supported by the STFC rolling grant STJ000434/1. MJP would like to thank the Mitchell foundation and Trinity College Cambridge for their generous support. PW thanks the STFC for support from the rolling grant awarded to King's College. DSB, MJP and PW would like to thank George Mitchell and Sheridan Lorenz for their generous hospitality at Cook's Branch.

\appendix

\section{Normalisation of generators}
\label{A}

In this appendix, we will derive an invariant scalar product which has
implicitly been used to construct the actions given in this paper.
Acting with  the  Cartan involution $I_c$ on the first
fundamental representation $l_1$ we can define  a new representation
$I_c(l_1)$  by
\begin{equation}
I_c(P_a)=-\bar P^a,\ I_c(Z^{ab})=-\bar Z_{ab},\ I_c(Z^{a_1\ldots
a_5})=-\bar Z_{a_1\ldots a_5}, \ldots
\label{(108)}
\end{equation}
where $\bar P^a$, $\bar Z_{ab}$, $\bar Z_{a_1\ldots a_5},\ldots $ are
elements of the representation $I_c(l_1)$.
\par
The Cartan involution $I_{c}$ takes negative root generators to
positive root
generators up to a sign in such a way as to preserve the algebra. A more
fundamental definition can be found in \cite{w42}, for example. The action of $I_{c}$ on
some of the $E_{11}$
generators is given by
\begin{equation}
I_c(K^a{}_b)= -K^b{}_a,\ I_c(R^{a_1a_2a_3})=-R_{a_1a_2a_3},\
I_c(R^{a_1\ldots a_6})=R_{a_1\ldots a_6}.
\label{(109)}
\end{equation}
The Cartan involution interchanges upper and lower indices, and
possibly involves a change of sign. Consistency of the commutation
rules under $I_{c}$ determines uniquely the sign.
Given equations \eqref{(108)} and \eqref{(109)} we can derive the
commutation relations
between $E_{11}$ and those of the  $I_c(l_1)$ representation. For
example,  acting with the Cartan involution on the commutator
$[R^{a_1a_2a_3}, P_b]=3\delta^{[a_1}_b Z^{a_2a_3]}$ we find that
\begin{equation}
[R_{a_1a_2a_3},
\bar P^b]=-3\delta_{[a_1}^b \bar Z_{a_2a_3]}.
\label{(110)}
\end{equation}
Using equations \eqref{(37b)}, \eqref{(38a)}, \eqref{(39c)} and \eqref{(39d)}, we find using similar arguments that
\begin{gather*}
[R_{a_1a_2a_3},\bar Z_{b_1b_2} ]= -\bar Z_{a_1a_2a_3
b_1b_2} \\
[K^a{}_b, \bar P^c ]= \delta_b^c\bar P^a-{1\over 2}
\delta_a^b \bar P^c,\\
[R^{a_1a_2a_3},\bar P^b]=0,  \\
[R^{a_1a_2a_3}, \bar Z_{b_1b_2} ]=-6\delta ^{[a_1a_2}_{b_1b_2}
\bar P^{a_3]}.
\end{gather*}
\par
Given any  element $A$ of the $l_1$ representation and any element $B
$ of the
$I_c(l_1)$ we can form an invariant scalar product denoted $(A,B)$; the
invariance means that
\begin{equation}
([X,A], \bar B)= -(A, [X,\bar B]),\quad X\in E_{11},\  A\in l_1, \
\bar B\in
\bar l_1.
\label{(112)}
\end{equation}
Taking $X=R^{a_1a_2a_3}$, $A=P_a$ and $B=\bar Z_{b_1b_2}$ we find using
equation \eqref{(112)} and equation \eqref{(110)} that
\begin{equation}
2\delta^{[a_1a_2}_{b_1b_2}(P_c, \bar  P^{a_3]})=
\delta^{[a_1}_c(Z^{a_1a_2]}, \bar Z_{b_1b_2}).
\label{(113)}
\end{equation}
In fact, choosing our normalisation and  using invariance under SL(11) we
must set
\begin{equation}
(P_c,\bar  P^{a})=\delta_c^a
\label{(114)},
\end{equation}
hence $(Z^{a_1a_2}, \bar Z_{b_1b_2})=2\delta ^{a_1a_2}_{b_1b_2}.$
Using similar arguments, and repeating the above result,  we find that
\begin{gather}
(P_b,\bar  P^{a})=\delta_b^a,\qquad
(Z^{a_1a_2}, \bar Z_{b_1b_2})=2\delta ^{a_1a_2}_{b_1b_2} , \qquad
(Z^{a_1\ldots a_5}, \bar Z_{b_1\ldots b_5})=5!\delta
^{a_1\ldots a_5}_{b_1\ldots b_5} , \notag \\
(Z^{a_1\ldots a_7, c}, \bar Z_{b_1\ldots b_7, d})=9 (7!) \delta
^{a_1\ldots a_7}_{b_1\ldots b_7} \delta^{c}_{d}.
\label{(115)}
\end{gather}
Let us write the scalar product for all generators  in the form
\begin{equation}
(L, \bar L)=N, \quad L\in l_1, \quad \bar L\in I_c(l_1)
\label{(116)}
\end{equation}
where $N$ is a diagonal matrix.
\par
We will now derive an equation for the object $M$, that we have used to
construct the Lagrangians,  in terms of the generalised vielbein $E$.
This
will involve the matrix $N$ just introduced. Let us first recall  the
technical steps given in section \ref{3} leading to the appearance of
the
generalised vielbein  in the non-linear realisation.
We can write the group element
$g_l$  in the form
$g_l=e^{z^T\cdot L'}$ where $L^{\prime}= CL$ and $C$ is a diagonal
matrix
which takes account of the possible normalisation factors.  The
Cartan form contains the terms
$g_l^{-1}d g_l= dz^T\cdot L^{\prime}$. Acting with the Cartan involution
we find that $I_c(g_l)= e^{\bar  L^{\prime}\cdot \bar z^T } $ and  so
$I_c(g_l^{-1}d g_l)=\bar  L^{\prime}\cdot d\bar z^T$. It is easy to see
that
\begin{equation}
(g^{-1}_ldg_l, I_c(g^{-1}_ldg_l))=dz^T \cdot C N C  \cdot d\bar z.
\label{(117)}
\end{equation}
\par
We take group element    $k\in E_{11}$  to act on the generators of the $l_1$ representation as $k^{-1}
L^{\prime} k= D(k) L^{\prime}$ and as a result the part of the
Cartan form that contains the generalised vielbein $E$ is given by
\begin{equation}
g_E^{-1}( g_l^{-1}d g_l) g_E= dz^T\cdot D(g_E)\cdot L^{\prime}\equiv
dz^T\cdot E \cdot L^{\prime}.
\label{(118)}
\end{equation}
Using equation \eqref{(118)} and \eqref{(60)} we find that
\begin{equation}
(I_c (g_E^{-1}))^{-1}g_E^{-1}( g_l^{-1}d g_l) g_E I_c(g_E^{-1})
= dz^T\cdot D(g_E)D(I_c(g_E^{-1}))\cdot L^{\prime}\equiv
dz^T\cdot M\cdot L^{\prime}.
\label{(119)}
\end{equation}
Let us now consider the object
\begin{equation}
((I_c (g_E^{-1}))^{-1}g_E^{-1}( g_l^{-1}d g_l) g_E I_c(g_E^{-1}),
I_c( g_l^{-1}d g_l))= dz^T\cdot M\cdot CNC \cdot d\bar z.
\label{(120)}
\end{equation}
Using the invariance of the scalar product \eqref{(112)}, which is
equivalent to
$$ (g_{0} A g_{0}^{-1}, \bar{B}) = (A, g_{0}^{-1} \bar B g_{0}),
\quad  A\in l_1, \bar B\in
\bar l_1, $$
where $g_{0}$ is an $E_{11}$ group element, we find that the object on the left-hand side of equation \eqref{(120)} is invariant under both the rigid and local
transformations  given in equation \eqref{(47)} and \eqref{(48)}.
Using again the invariance of the scalar product and equation \eqref{(116)} we
can evaluate this object to find that
\begin{align}
dz^T\cdot M\cdot CNC \cdot d\bar z
&=(g_E^{-1}( g_l^{-1}d g_l) g_E, I_c(g_E^{-1}) I_c(
g_l^{-1}d g_l)(I_c (g_E^{-1}))^{-1})\notag \\
&= (g_E^{-1}( g_l^{-1}d g_l) g_E, I_c(g_E^{-1}
g_l^{-1} d g_lg_E) )\notag \\
&=(dz^T\cdot E \cdot L^{\prime},( {\bar L}^\prime)^T\cdot E^T \cdot
d\bar z) \notag \\
&=dz^T \cdot EC NC \cdot E^T \cdot d\bar z.
\label{(121)}
\end{align}
Hence we find that
$MCNC= ECNCE^T$.
We will choose $C$ so that $CNC=I$ and then
\begin{equation}
M=EE^T
\label{(122)}
\end{equation}
This choice also implies that
\begin{equation}
(L^\prime, \bar L^\prime )=I
\label{(123)}
\end{equation}
and equation \eqref{(117)} becomes
\begin{equation}
(g^{-1}_ldg_l, I_c(g^{-1}_ldg_l))=dx^a d\bar x_a+ dx^{ab}d\bar
x_{ab}=\dots.
\label{(124)}
\end{equation}

In the case of the SL(5) duality group found in dimension 4, $C$ is
the diagonal matrix with diagonal entries
$$(1, {1\over{\sqrt{2}}} ),$$
so the group element $g_l$ takes the form $$e^{x^iP_i + {1\over \sqrt
2} {x_{ij} Z^{ij} }}$$ in equation \eqref{(75)}.
In dimension 5, the dual of the generator $Z^{a_1 \dots a_5}$  has
been used. The normalisation of $W= {1\over 5!} \epsilon_{a_1 \dots
a_5} Z^{a_1 \dots a_5}$ can easily be found from equation \eqref{(115)},
$$(W, \bar{W})= 1, $$ so in this case $C$ has diagonal entries $$(1,
{1\over{\sqrt{2}}}, 1 ).$$
Similarly, in dimension 6, the dual of the $Z^{a_1 \dots a_5}$ is $W_
{a}= {1\over 5!} \epsilon_{a b_1 \dots b_5} Z^{b_1 \dots b_5},$ which
from equation \eqref{(115)} has the normalisation
$$(W_{a}, \bar{W}^{b})= \delta^{b}_{a},$$ so in six dimensions $C$
also has diagonal entries $$(1, {1\over{\sqrt{2}}}, 1 ).$$
In seven dimensions, we have used the Hodge dual of two of the
translation generators,
$$W_{ab}= {1\over 5!} \epsilon_{a b c_1 \dots c_5} Z^{c_1 \dots c_5},
\; W^{a} = {1\over 7!} \epsilon_{b_1 \dots b_7} Z^{b_1 \dots b_7, a}.$$
The normalisation of these generators is found to be
$$(W_{a b}, \bar{W}^{cd})= 2 \delta^{cd}_{ab}, \; (W^{a}, \bar{W}_{b})
= 9 \delta^{a}_{b}. $$
Therefore, in the case of seven dimensions $C$ has diagonal entries
$$(1,{1\over{\sqrt{2}}},{1\over{\sqrt{2}}}, {1\over3}).$$

\section{Rescaling of the generalised metric}
\label{B}
In this appendix, we show that rescaling a generalised metric by its
determinant gives a generalised metric that also reproduces the
dynamical theory. There are some important caveats that will be
explained. Assume that a generalised metric, $M,$ reproduces the
dynamics, when the fields only depend on the ordinary space
coordinates and not on the extra generalised coordinates,
\begin{multline}
L = c_{1}\, M^{MN} (\partial_M M^{KL})( \partial_N
M_{KL} ) + c_{2}\, M^{MN} (\partial_N M^{KL}) (\partial_L M_{MK}) \\
+ c_{3} \,   M^{MN} M^{PQ}(M^{RS} \partial_P M_{RS}) (\partial_M M_
{NQ})
+ c_{4}\, M^{MN} (M^{KL} \partial_M M_{KL})(M^{RS} \partial_N M_{RS}), 
\label{(125)}
\end{multline}
where $c_{1}, \dots, c_{4}$ are known real numbers. We also require
that the determinant of $M$ is related to the determinant of the
space metric $g$,
\begin{equation}
   \det M = g^{a},
\label{detMa}
\end{equation}
for some real constant $a.$ This is required by gauge-invariance of
the theory under gauge transformations of the potential 3-form and 6-form.

Consider rescaling of the generalised metric $M$ by its determinant,
or equivalently $g,$
\begin{equation}
\tilde{M} = g^{\alpha} M,
\label{alpha}
\end{equation}
where $\alpha$ is a real number. Therefore, $\tilde{M}^{-1} = g^{-
\alpha} M^{-1}$ and so
\begin{align*}
\tilde{M}^{MN} (\partial_M \tilde{M}^{KL})( \partial_N
\tilde{M}_{KL} ) =&
g^{-\alpha} M^{MN} (\partial_M M^{KL})( \partial_N
M_{KL} ) \\ & \qquad - {{\alpha}\over{a}} \left(2+ {{\alpha}\over{a}}
D \right) g^{-\alpha} M^{MN} (M^{KL} \partial_M M_{KL})(M^{RS}
\partial_N M_{RS}),
\end{align*}
where $D$ is the dimension of generalised space, and we have used
\begin{equation}
   M^{KL} \partial_M M_{KL} = {{\partial_{M} (\det M)}\over{\det M}}=
a {{\partial_{M} g}\over{g}}.
\label{detMa2}
\end{equation}
Similarly,
\begin{align*}
\tilde{M}^{MN} (\partial_N \tilde{M}^{KL}) (\partial_L \tilde{M}_{MK})
=& g^{-\alpha} M^{MN} (\partial_N M^{KL}) (\partial_L M_{MK}) \\ &
\qquad - {{2 \alpha}\over{a}}g^{-\alpha} M^{MN} M^{PQ}(M^{RS}
\partial_P M_{RS}) (\partial_M M_{NQ}) \\ & \qquad \qquad- {{\alpha^2}
\over{a^2}} g^{-\alpha} M^{MN} (M^{KL} \partial_M M_{KL})(M^{RS}
\partial_N M_{RS}),
\end{align*}
\begin{align*}
\tilde{M}^{MN} (\tilde{M}^{KL} \partial_M  \tilde{M}_{KL})(\tilde{M}^
{RS} \partial_N \tilde{M}_{RS}) =  \left( {{\alpha}\over{a}} D +1
\right)^2 g^{-\alpha} M^{MN} (M^{KL} \partial_M M_{KL})(M^{RS}
\partial_N M_{RS}),
\end{align*}
\begin{align*}
\tilde{M}^{MN} \tilde{M}^{PQ}  (\tilde{M}^{RS} \partial_P \tilde{M}_
{RS}) (\partial_M \tilde{M}_{NQ})
&= \left( {{\alpha}\over{a}} D +1 \right) g^{-\alpha} M^{MN} M^{PQ}(M^
{RS} \partial_P M_{RS}) (\partial_M M_{NQ})  \\ & + {{\alpha}\over
{a}} \left( {{\alpha}\over{a}} D +1 \right) g^{-\alpha} M^{MN} (M^
{KL} \partial_M M_{KL})(M^{RS} \partial_N M_{RS}).
\end{align*}
Hence, as long as $$\left( {{\alpha}\over{a}} D +1 \right)\neq0,$$
the rescaled generalised metric also reproduces the action, but with
different coefficients for last two terms, i.e.\ $c_1$ and $c_2$ will
have the same value, but the value of the constants $c_3$ and $c_4$
will change.

The case where $$\left( {{\alpha}\over{a}} D +1 \right)$$ vanishes
actually corresponds to the case where the generalised metric is
derived from the duality group algebra. For example for the SL(5)
duality group, let $M$ denote the generalised metric \begin{equation}
M_{KL} =
\begin{pmatrix} g_{\mu \nu}+{{1}\over{2}} C_{\mu}{}^{ij} C_{\nu ij} &
-{{1}\over{\sqrt{2}}} C_{\mu}{}^{\nu_1 \nu_2 } \\ -{{1}\over{\sqrt
{2}}} C^{\mu_1 \mu_2}{}_{\nu}  & g^{\mu_1 \mu_2, \nu_1 \nu_2}
\end{pmatrix},
\end{equation}
then the generalised metric derived from the SL(5) motion group is $
\tilde{M}=g^{1/5} M,$ equation \eqref{(22)} in section \ref{2}
\footnote{The $M$ in section \ref{2} is $\tilde{M}$ here.}. From
equation \eqref{alpha}, $\alpha= 1/5,$ and from equation \eqref
{detMa}, or \eqref{detMa2}, $a=-2.$ The dimension of the generalised
space, $D,$ is 10. Hence $$\left( {{\alpha}\over{a}} D +1 \right)=0.$$
In contrast, for the generalised metric from the non-linear
realisation of $E_{11} \ltimes l_1,$ equation \eqref{(81)}, the
corresponding values of $\alpha, a $ and $D$ are $-1/2, -2$ and $10,$
so $$\left( {{\alpha}\over{a}} D +1 \right) \neq 0.$$

It can easily be checked that the above statement is also true for
the SO(5,5), $E_{6}$ and $E_{7}$ duality groups. In all these cases,
let $M$ denote the generalised metric with no factor of $\det g$ in
its top-left entry, and $\tilde{M}$ be the generalised metric from
the non-realisation of the duality motion group. Then as can be seen
from table \ref{table3}, $$\left( {{\alpha}\over{a}} D +1 \right) = 0$$
in all these cases. Therefore, the generalised metric constructed
from the duality group cannot be used to reproduce the dynamics.
However, if the generalised metrics come from the non-linear
realisation of larger groups such as $E_{9}, E_{10}$ or $E_{11},$
then the value of $\alpha$ is different and the generalised metric
can be used to construct the dynamics. The particular advantage of $E_{11}$ is that it not only solves the above problem, but that it also results in the correct overall measure.

\begin{table}[htbp]
\centering
\begin{tabular}{ | c | c | c | c | c | }
\hline
& SL(5) & SO(5,5) & $E_{6}$ & $E_{7}$ \\ \hline
$\alpha$ & 1/5 & 1/4 & 1/3 & 1/2 \\
$a$ &  -2 & -4 & -9 & -28 \\
$D$ & 10 & 16 & 27 & 56 \\ \hline
\end{tabular}
\caption{The values of $\alpha, a $ and $D$ for the duality groups
considered in this paper.}
\label{table3}
\end{table}

\section{$E_7$ motion group from Cartan's representation}
\label{C}

Here, we will briefly review Cartan's 56-dimensional representation
of $E_7$ \cite{cartanthese, cartan, freudenthal} and use it to find
the algebra of the $E_{7}$ motion group\footnote{See also appendix B
of \cite{so(8)} for a complementary account of $E_{7}$}. We show that
the truncation of the $E_{11} \ltimes l_{1}$ at lowest level to seven
dimensions gives the algebra of the $E_{7}$ motion group.

We will consider the representation of the exceptional Lie group $E_{7}$ on a 56-dimensional space parametrised by bivectors, $x^{IJ},$
and 2-form $y_{IJ},$ where $I,J$ run from 1 to 8. The infinitesimal
transformations of these under $E_{7}$ are
\begin{gather}
x^{IJ} \rightarrow x^{IJ} + \Lambda^{I}_{\;\,K} x^{KJ} + \Lambda^{J}_
{\;\,K} x^{IK} + \Sigma^{IJKL} y_{KL} \label{(126)}\\
y_{IJ} \rightarrow y_{IJ} - \Lambda^{K}{}_{I} y_{KJ} - \Lambda^{K}{}_
{J} y_{IK} + \Sigma_{IJKL} x^{KL}, \label{(127)}
\end{gather}
where $\Lambda^{I}_{\;\,I}=0,$ and $$ \Sigma^{IJKL} = {{1}\over{4!}}
\epsilon^{IJKLMNPQ} \Sigma_{MNPQ}.$$
The $\Lambda$ and $\Sigma$ parametrise the infinitesimal $E_{7}$
transformations.

To find the commutation relations of the motion group, we denote an
$E_7$ motion group transformation by
\begin{equation}
U(\Lambda, \Sigma; a, b)= \textup{e}^{\Lambda^{J}_{\;\,I} M^{I}_{\;
\,J} + \Sigma^{IJKL} V_{IJKL} + a^{IJ} X_{IJ} + b_{IJ} Y^{IJ}},
\label{(128)}
\end{equation}
where $X^{IJ} y_{KL} =0,$ and $Y_{IJ} x^{KL} =0.$ The generators $M^{I}_{\;\,J}$ and  $V_{IJKL}$ generate $E_{7}$ transformations
parametrised by $\Lambda^{I}_{\;\,J}$ and $\Sigma^{IJKL},$
respectively, and $X^{IJ}$ generates translations in the $x^{IJ}$
directions, while $Y_{IJ}$ generates translations in the $y_{IJ}$
directions. The transformation of $x^{IJ}$ and $y_{IJ}$ under the $E_{7}$ part of $U$ is given, to first order, in equations \eqref{(126)}
and \eqref{(127)}, respectively.

The commutator of two transformations can be used to calculate the
commutation relations of the generators. To this end, we calculate
the commutator of two transformations on $x^{IJ}$ and $y_{IJ}$ to
second order in the infinitesimal parameters
\begin{align}
&[ \tilde{U}(\tilde{\Lambda}, \tilde{\Sigma}; \tilde{a}, \tilde{b}),
U(\Lambda, \Sigma; a, b) ] x^{IJ} \notag  \\
=& \left(  [ \tilde{\Lambda}, \Lambda]^{I}_{\;\,K} - {{1}\over{3}}
\Theta^{I}_{\;\,K} \right) x^{KJ}  + \left(  [ \tilde{\Lambda},
\Lambda]^{J}_{\;\,K} - {{1}\over{3}} \Theta^{J}_{\;\,K} \right) x^
{IK} \notag \\
& - 4 \left( \tilde{\Lambda}^{[I}{}_{K} \Sigma^{JMN]K} - \Lambda^{[I}{}_
{K} \tilde{\Sigma}^{JMN]K} \right) y_{MN} + \tilde{\Lambda}^{I}{}_{K} a^{KJ} - \Lambda^{I}{}_{K} \tilde{a}^{KJ}\notag \\
& \qquad  + \tilde{\Lambda}^{J}_{\;\,K} a^{IK} - \Lambda^{J}_{\;\,K}
\tilde{a}^{IK} + \tilde{\Sigma}^{IJKL} b_{KL} - \Sigma^{IJKL} \tilde
{b}_{KL},
\label{(129)}
\end{align}
where $$ \Theta^{I}_{\;\,J}=  \tilde{\Sigma}^{IKLM} \Sigma_{KLMJ} - \Sigma^{IKLM} \tilde{\Sigma}_{KLMJ}.$$ There is a similar expression for the commutator of two transformations acting on $y_{IJ}$
\begin{align}
&[ \tilde{U}(\tilde{\Lambda}, \tilde{\Sigma}; \tilde{a}, \tilde{b}),
U(\Lambda, \Sigma; a, b) ] y_{IJ} \notag  \\
=& - \left(  [ \tilde{\Lambda}, \Lambda]^{K}{}_{I} - {{1}\over{3}}
\Theta^{K}_{\;\,I} \right) y_{KJ} - \left(  [ \tilde{\Lambda},
\Lambda]^{K}{}_{J} - {{1}\over{3}} \Theta^{K}_{\;\,J} \right) y_
{IK} \notag \\
&\; + 4 \left( \tilde{\Lambda}^{K}{}_{[I} \Sigma_{JMN]K} - \Lambda^
{K}{}_{[I} \tilde{\Sigma}_{JMN]K} \right) x^{MN} + \Lambda^{K}{}_{I} \tilde{b}_{KJ} - \tilde{\Lambda}^{K}{}_{I} b_{KJ}\notag \\
& \qquad   + \Lambda^{K}{}_{J} \tilde{b}_{IK}  - \tilde{\Lambda}^{K}{}_{J} b_{IK}+ \tilde{\Sigma}_{IJKL} a^{KL} - \Sigma_{IJKL} \tilde
{a}^{KL}.
\label{(130)}
\end{align}
In the above equations we have used the identity
$$ \tilde{\Sigma}^{IJKL} \Sigma_{KLMN} - \Sigma^{IJKL} \tilde{\Sigma}_
{KLMN} = - {{2}\over{3}} \delta^{[I}_{[M} \Theta^{J]}_{\;\,N]},$$
which can be proved by Hodge dualising $\tilde{\Sigma}$ and $\Sigma$
and then contracting the epsilon tensors, and expanding out the
antisymmetrisations in the resulting Kronecker delta symbols.

Hence, from the above equations, \eqref{(129)} and \eqref{(130)}, we
deduce that the commutator of two transformations $\tilde{U}$ and $U$
is an infinitesimal transformation, as it must be from Lie theory,
and the transformation can be written
\begin{align}
&[ \tilde{U},  U ] \notag \\
&=  \left([ \tilde{\Lambda}, \Lambda]- {{1}\over{3}} \Theta\right)^{J}{}_{I} M^{I}_{\;\,J} + 4 \left( \tilde{\Lambda}^{I}{}_{M} \Sigma^
{MJKL} - \Lambda^{I}{}_{M} \tilde{\Sigma}^{MJKL} \right) V_{IJKL}
\notag \\
& \qquad\quad  + \left( 2 \tilde{\Lambda}^{I}{}_{K} a^{KJ} - 2
\Lambda^{I}{}_{K} \tilde{a}^{KJ}  + \tilde{\Sigma}^{IJKL} b_{KL} -
\Sigma^{IJKL} \tilde{b}_{KL} \right) X_{IJ}  \notag \\
& \qquad\qquad\qquad\quad   + \left( 2 \Lambda^{K}{}_{I} \tilde{b}_
{KJ} - 2 \tilde{\Lambda}^{K}{}_{I} b_{KJ}+ \tilde{\Sigma}_{IJKL} a^
{KL} - \Sigma_{IJKL} \tilde{a}^{KL} \right) Y^{IJ}.
\label{(131)}
\end{align}

Now using the above equation we can find the commutation relations.
For example, from the above equation
\begin{equation}
[ \tilde{U} (\tilde{\Lambda}, 0; 0,0) ,  U(\Lambda, 0; 0,0) ] =
\textup{e}^{[ \tilde{\Lambda}, \Lambda]^{J}_{\;\,I} M^{I}_{\;\,J}}.
\label{(132)}
\end{equation}
But the $\tilde{U}$ and $U$ can also be written using exponentials,
equation \eqref{(128)}, so the commutator of the two transformations
can also be written as
\begin{align}
[\tilde{U} (\tilde{\Lambda}, 0; 0,0) ,  U(\Lambda, 0; 0,0) ] &=
\textup{e}^{\tilde{\Lambda}^{J}_{\;\,I} M^{I}_{\;\,J}} \textup{e}^
{\Lambda^{L}_{\;\,K} M^{K}_{\;\,L}} - \textup{e}^{\Lambda^{L}_{\;\,K}
M^{K}_{\;\,L}} \textup{e}^{\tilde{\Lambda}^{J}_{\;\,I} M^{I}_{\;
\,J}}, \notag \\
&= \tilde{\Lambda}^{J}_{\;\,I} \Lambda^{L}_{\;\,K} [M^{I}_{\;\,J}, M^
{K}_{\;\,L}],
\label{(133)}
\end{align}
using the Baker-Campbell-Hausdorff formula $$ \textup{e}^X \textup{e}
^Y = \textup{e}^{X+Y + {{1}\over{2}} [X,Y] \dots}.$$
Comparing equations \eqref{(132)}and \eqref{(133)}, we deduce that
\begin{equation}
   [M^{I}_{\;\,J}, M^{K}_{\;\,L}] = \delta^{I}_{L} M^{K}_{\;\,J} -
\delta^{K}_{J} M^{I}_{\;\,L}.
\label{(134)}
\end{equation}

The other commutation relations can be found using the same method
and are listed below:
\begin{gather}
   [M^{I}_{\;\,J}, V_{ABCD}] = 4 \, \delta^{I}_{[A} V_{|J|BCD]} - {{1}
\over{2}} \, \delta^{I}_{J} V_{ABCD}, \label{(135)} \\  [V_{ABCD}, V_
{EFGH}] = - {{1}\over{72}} \, \left( \delta^{J}_{[A} \epsilon_{BCD]
EFGHI} - \delta^{J}_{[E} \epsilon_{FGH]ABCDI} \right) M^{I}_{\;\,J}, \\
[M^{I}_{\;\,J}, X_{KL}] = 2 \, \delta^{I}_{[K} X_{|J|L]} - {{1}\over
{4}} \, \delta^{I}_{J} X_{KL},  \\ [M^{I}_{\;\,J}, Y^{KL}] = - 2 \,
\delta^{[K}_{J} Y^{|I|L]} + {{1}\over{4}} \, \delta^{I}_{J} Y^{KL}, \\
[V_{ABCD}, X_{IJ}] = {{1}\over{4!}} \, \epsilon_{ABCDIJKL} Y^{KL},
\quad  [V_{ABCD}, Y^{IJ}] = \delta^{KL}_{[AB} X_{CD]}. \label{(139)}
\end{gather}
These are the commutation relations of SL(8) decomposition of the
algebra of the $E_7$ motion group. The uppercase Latin indices are in
fact SL(8) indices, which is why they run from 1 to 8. We are,
however, interested in the SL(7) decomposition of the algebra of the
$E_7$ motion group. This is because the $E_7$ duality appears upon
reduction on a 7-torus, so we will make the duality act along these
seven spatial directions.

It is not difficult to decompose SL(8) representations in terms of SL(7) representations. We let $I=(i,8),$ where lowercase Latin letters
are SL(7) indices that run from 1 to 7, and we define
\begin{gather}
M^{i}_{\;\;j} = - \tilde{K}^{i}_{\;\;j} + {{1}\over{6}} \delta^{i}_
{j} D, \\  M^{8}_{\;\;i}={{2}\over{6!}} \epsilon_{i k_1 \dots k_6} R^
{a_1 \dots a_6}, \qquad M^{i}_{\;\;8}= - {{2}\over{6!}} \epsilon^{i
k_1 \dots k_6} R_{k_1 \dots k_6}, \label{(141)} \\
V_{ijk8} = {{1}\over{12}} R_{ijk}, \quad V_{ijkl} = {{1}\over{72}}
\epsilon_{ijklmnp} R^{mnp}, \\
X_{i8}= {{1}\over{\sqrt{2}}} P_{i}, \quad X_{ij}= {{1}\over{\sqrt
{2}}} W_{ij}, \quad Y^{i8}={{1}\over{3\sqrt{2}}} W^{i}, \quad Y^{ij}=
{{1}\over{\sqrt{2}}} Z^{ij},
\end{gather}
where $D= \sum_{i} \tilde{K}^{i}_{\;\,j}.$ The normalisation has been
chosen to match the normalisation of the $E_{11}\ltimes l_{1}$
generators in section \ref{3}. In particular, the coefficient of $D$
in the relation between $M^{i}_{\;\;j}$ and $\tilde{K}^{i}_{\;\;j},$
the first equation in the set of equations \eqref{(141)}, has been
chosen so that the commutator of $\tilde{K}^{i}_{\;\;j}$ and $R^
{ijk}, R_{ijk}, R^{i_1 \dots i_6}$ and $R_{i_1 \dots i_6}$ has no
trace term.

The commutation relations for the SL(7) decomposition of the $E_7$
motion group are found by inserting the decomposed generators into
the commutation relations \eqref{(135)}--\eqref{(139)}. Whereupon,
the $E_7$ commutation relations are
\begin{gather}
[ \tilde{K}^{i}_{\;j} , \tilde{K}^{k}_{\;l} ] = \delta^{k}_{j} \tilde
{K}^{i}_{\;l} - \delta^{i}_{l} \tilde{K}^{k}_{\;j}, \\
[ \tilde{K}^{i}_{\;j} , R_{klm}] = -3 \, \delta^{i}_{[k} R_{|j|lm]},
\qquad
[ \tilde{K}^{i}_{\;j} , R^{klm}] = 3 \, \delta^{[k}_{j} R^{|i|lm]}, \\
[ \tilde{K}^{i}_{\;j} , R_{k_1 \dots k_6}] = - 6 \, \delta^{i}_{[k_1}
R_{|j|k_2 \dots k_6]},  \qquad
[ \tilde{K}^{i}_{\;j} , R^{k_1 \dots k_6}] = 6 \, \delta^{[k_1}_{j} R^
{|i|k_2 \dots k_6]}, \\
[ R_{i_1 \dots i_3} , R_{j_1 \dots j_3}] = 2R_{i_1 \dots i_3j_1 \dots
j_3}, \qquad [ R^{i_1 \dots i_3} , R^{j_1 \dots j_3}] = 2 R^{i_1
\dots i_3 j_1 \dots j_3}, \\
[ R_{i_1 \dots i_3} , R^{j_1 \dots j_6}] =  60 \, \delta_{i_1 \dots
i_3}^{[j_1 \dots j_3} R^{j_4 \dots j_6]}, \quad [ R^{i_1 \dots i_3} ,
R_{j_1 \dots j_6}] = -60 \, \delta^{i_1 \dots i_3}_{[j_1 \dots j_3} R_
{j_4 \dots j_6]}, \\
[ R^{i_1 \dots i_3} , R_{j_1 \dots j_3}] = 18 \, \delta^{[i_1 i_2}_
{[j_1 j_2} \tilde{K}^{i_3]}_{\;\; j_3]} -2 \, \delta^{i_1 \dots i_3}_
{j_1 \dots j_3} D,  \\
[ R^{i_1 \dots i_6} , R_{j_1 \dots j_6}] = -5!3.3 \, \delta^{[i_1
\dots i_5}_{[j_1 \dots j_5} \tilde{K}^{i_6]}_{\;\;j_6]} +5 ! \,
\delta^{i_1 \dots i_6}_{j_1 \dots j_6} D.
\end{gather}
Furthermore, the commutation relations of the $E_7$ generators with
the translation generators are
\begin{gather}
[ \tilde{K}^{i}_{\;j} , P_{k} ] = - \delta^{i}_{k} P_{j} - {{1}\over
{2}} \delta^{i}_{j} P_{k}, \qquad
[ \tilde{K}^{i}_{\;j} , Z^{kl} ] = 2 \, \delta^{[k}_{j} Z^{|i|l]} -
{{1}\over{2}} \delta^{i}_{j} Z^{kl},  \\
[ \tilde{K}^{i}_{\;j} , W_{kl} ] = - 2 \, \delta^{i}_{[k} W_{|j|l]} +
{{1}\over{2}} \delta^{i}_{j} W_{ij}, \qquad
[ \tilde{K}^{i}_{\;j} , W^{k} ] =  \delta^{k}_{j} W^{i} + {{1}\over
{2}} \delta^{i}_{j} W^{k}, \\
[ R_{ijk} , P_{k} ] = 0, \qquad
[ R_{ijk} , Z^{mn} ] = 3! \, \delta^{mn}_{[ij} P_{k]}, \\
[ R_{ijk} , W_{mn} ] = {{1}\over{2}} \epsilon_{ijkmnpq} Z^{pq}, \qquad
[ R_{ijk} , W^{l} ] = 9 \delta^{l}_{[i} W_{jk]}, \\
[ R^{ijk} , P_{l} ] = 3 \, \delta^{[i}_{l} Z^{jk]}, \qquad
[ R^{ijk} , Z^{mn} ] = {{1}\over{2}} \epsilon^{ijkmnpq} W_{pq}, \\
[ R^{ijk} , W_{mn} ] = 2 \, \delta^{[ij}_{mn} W^{k]}, \qquad
[ R^{ijk} , W^{l} ] = 0, \\
[ R_{i_1 \dots i_6}, P_{j}] = 0, \qquad
[ R_{i_1 \dots i_6}, Z^{kl}] = 0, \\
[ R_{i_1 \dots i_6}, W_{kl}] = \epsilon_{j i_1 \dots i_6} \delta^{j}_
{[k} P_{l]}, \qquad
[ R_{i_1 \dots i_6}, W^{k}] = -{{3}\over{2}} \epsilon_{j i_1 \dots
i_6} Z^{jk}, \\
[ R^{i_1 \dots i_6}, P_{k}] = -{{1}\over{2}} \epsilon^{j i_1 \dots
i_6} W_{jk}, \qquad
[ R^{i_1 \dots i_6}, Z^{kl}] = {{1}\over{3}} \epsilon^{i_1 \dots i_6
[k} W^{l]}.
\end{gather}
The $E_{7}$ generators $\tilde{K}^{i}_{\;j}, R^{ijk}, R_{ijk},R^{i_1
\dots i_6},R_{i_1 \dots i_6} $ and the generalised translation
generators $ P_{i}, Z^{ij}, Z^{i_1 \dots i_5}, Z^{i_1 \dots i_7,j}$
can be exactly matched to the corresponding generators in section \ref{7}, which were derived from the $E_{11} \ltimes l_{1}$ algebra.

\end{document}